\documentclass[aps,twocolumn,prd,showpacs,showkeys,preprintnumbers,superscriptaddress,nobibnotes,floatfix,longbibliography,nofootinbib]{revtex4-1}

\pdfoutput=1
\usepackage{amsmath}
\usepackage{amsfonts}
\usepackage{amssymb}
\usepackage{mathrsfs}
\usepackage{graphicx}
\usepackage{xcolor}
\usepackage{longtable}
\usepackage[hidelinks]{hyperref}
\usepackage{multirow}
\usepackage{slashed}
\usepackage{bm}
\usepackage[capitalise]{cleveref}
\usepackage[normalem]{ulem}
\usepackage{array}

\newcommand*\DAlambert{\mathop{}\!\mathbin\Box}

\newcommand{\uvec}[1]{\hat{#1}}
\newcommand{\mAp}{m_{A'}}
\newcommand{\ie}{{\it i.e.}~}
\newcommand{\eg}{{\it e.g.}~}

\begin{document}

\title{Listening for Dark Photon Radio from the Galactic Centre}

\author{Edward Hardy}
\email{edward.hardy@liverpool.ac.uk}
\affiliation{Department of Mathematical Sciences, University of Liverpool, Liverpool, L69 7ZL, United Kingdom}
\author{Ningqiang Song}
\email{songnq@itp.ac.cn}
\affiliation{Department of Mathematical Sciences, University of Liverpool, Liverpool, L69 7ZL, United Kingdom}
\affiliation{Institute of Theoretical Physics, Chinese Academy of Sciences, Beijing, 100190, China}

\date{\today}

\begin{abstract}
Dark photon dark matter that has a kinetic mixing with the Standard Model photon can resonantly convert in environments where its mass $m_{A'}$ coincides with the plasma frequency. We show that such conversion in neutron stars or accreting white dwarfs in the galactic centre can lead to detectable radio signals. Depending on the dark matter spatial distribution, future radio telescopes could be sensitive to values of the kinetic mixing parameter that exceed current constraints by orders of magnitude for $m_{A'} \in \left(6\times 10^{-6},7\times 10^{-4}\right)$~eV.
\end{abstract}

\maketitle

\section{Introduction}

New light bosons, including axions and dark photons (DPs), are well-motivated extensions of the Standard Model (SM) that naturally arise in string theory compactifications~\cite{Svrcek:2006yi,Abel:2006qt,Abel:2008ai,Arvanitaki:2009fg,Goodsell:2009xc}. DPs are the vector bosons of extra U(1) gauge factors and can couple to the SM in several ways, perhaps most simply via kinetic mixing~\cite{Holdom:1985ag,Dienes:1996zr,Abel:2003ue}. A massive, sufficiently long-lived, DP might constitute dark matter; a cold relic population can be produced by numerous different mechanisms, \eg~\cite{Arias:2012az,Graham:2015rva,Dror:2018pdh,Co:2018lka,Bastero-Gil:2018uel,Agrawal:2018vin,Ema:2019yrd,Nakayama:2019rhg,Alonso-Alvarez:2019ixv,Nakai:2020cfw} (see however \cite{East:2022rsi} for complications). 
The DP dark matter parameter space has been explored experimentally by haloscopes~\cite{Godfrey:2021tvs,Nguyen:2019xuh,ADMX:2001nej,ADMX:2009iij,ADMX:2018gho,ADMX:2018ogs,ADMX:2019uok,Lee:2020cfj,HAYSTAC:2018rwy,HAYSTAC:2020kwv,Dixit:2020ymh,Alesini:2020vny,Cervantes:2022yzp,DOSUE-RR:2022ise,An:2022hhb,Knirck:2018ojz,Tomita:2020usq,Ramanathan:2022egk} as well as other approaches~\cite{Ehret:2010mh,Betz:2013dza,williams1971new,Caputo:2021eaa,An:2022hhb}. Additionally, the distortion of the cosmic microwave background (CMB) spectrum by conversion between DPs and photons leads to strong constraints~\cite{Jaeckel:2008fi,Mirizzi:2009iz,Arias:2012az,McDermott:2019lch} as does anomalous energy transfer in stars~\cite{raffelt1996stars,An:2013yfc,An:2013yua,Hardy:2016kme,An:2020bxd,Redondo:2013lna,Hong:2020bxo}.

Dark matter axions can convert to photons in the magnetosphere of neutron stars (NSs). Searches for the resulting radio waves could cover large parts of parameter space~\cite{Hook:2018iia,Huang:2018lxq,Safdi:2018oeu,Witte:2021arp,Millar:2021gzs,Battye:2021xvt,Battye:2021yue,Wang:2021hfb,Foster:2022fxn,Battye:2021xvt,Witte:2022cjj} and observations of the galactic centre (GC) already lead to interesting limits~\cite{Foster:2022fxn}. Kinetically mixed DPs can also efficiently convert to photons in environments where the DP mass $\mAp$ is approximately equal to the plasma frequency $\omega_p=\sqrt{4\pi\alpha n_e/m_e}$, where $n_e$ is the free electron number density and $\alpha$ is the fine structure constant. It has been suggested that such conversion in the solar corona could lead to observable signals from DPs with mass between $4\times 10^{-8}$~eV and $4\times 10^{-6}$~eV~\cite{An:2020jmf}. 

\begin{figure*}
    \centering
    \includegraphics[width=0.48\textwidth]{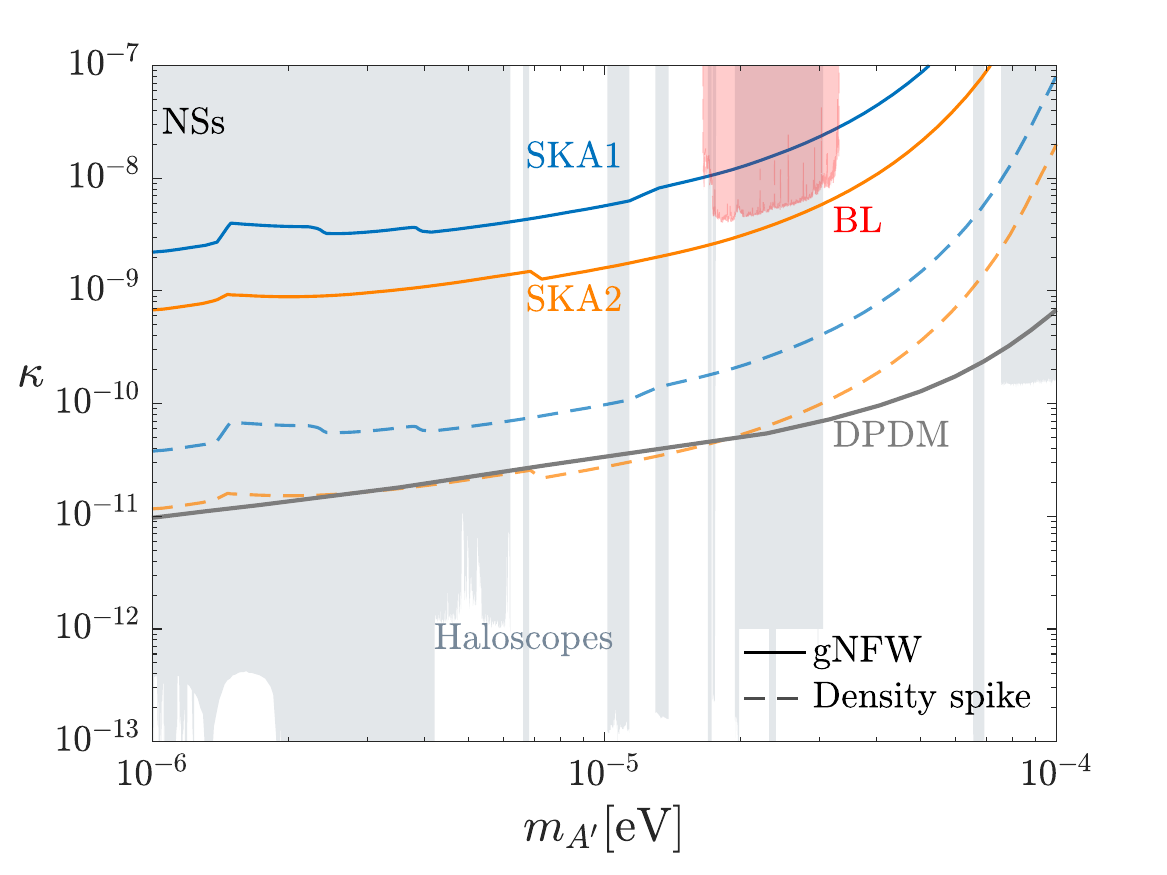}
    \includegraphics[width=0.48\textwidth]{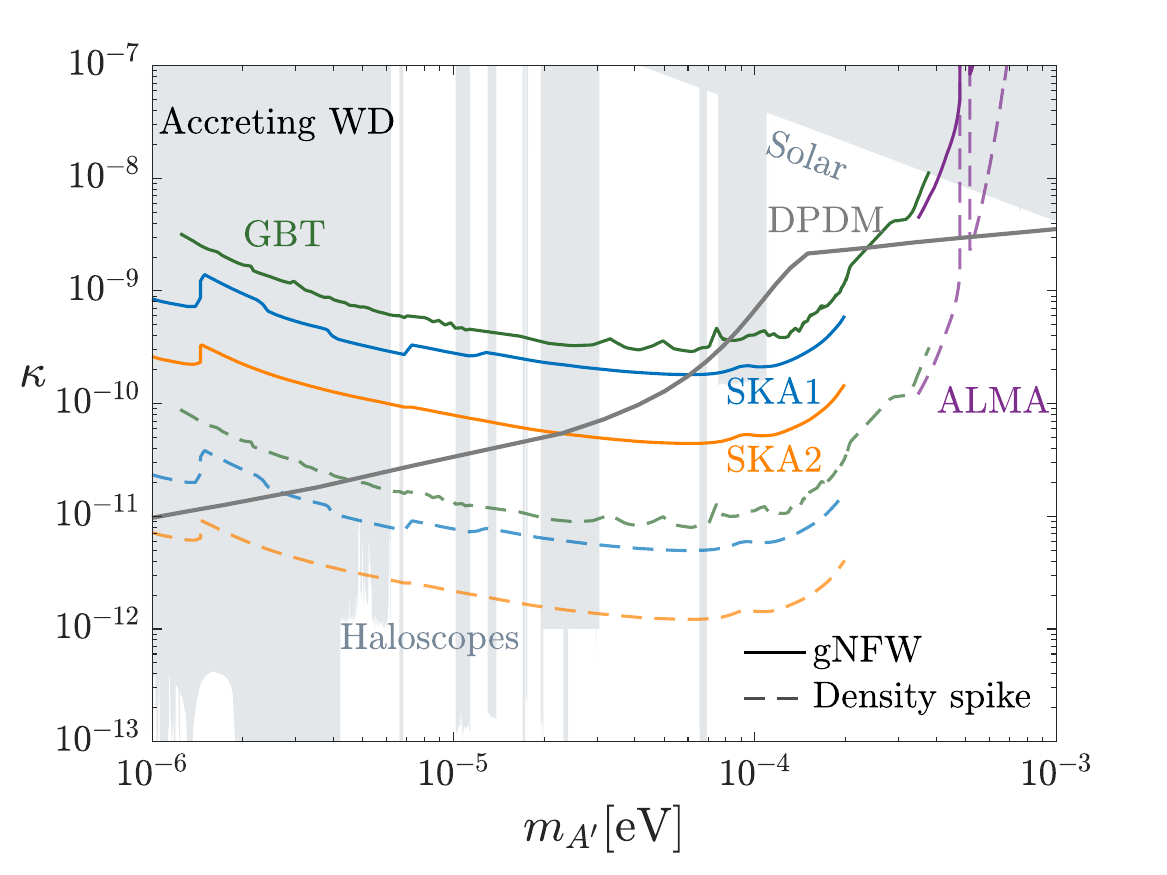} \caption{
Sensitivity of radio telescopes to dark photon dark matter from 100 hours of observation of the cumulative signal from neutron stars  within 3~pc of the galactic centre (NSs, left) and the signal from an individual accreting white dwarf  0.3~pc from the galactic centre (WD, right). Coloured lines give the projected sensitivities of ALMA, GBT, SKA1, and SKA2. The red shaded region is our constraint derived from the Breakthrough Listen (BL) project~\cite{Foster:2022fxn}. For each telescope we plot two lines: the solid lines assume a gNFW dark matter profile and the dashed lines assume a density spike near the galactic centre  (the BL limit is only visible with a density spike). The grey line depicts the cosmological constraint on dark photon dark matter (DPDM) from Arias et al~\cite{Arias:2012az}. Overlaid shaded regions are limits from haloscopes~\cite{Godfrey:2021tvs,Nguyen:2019xuh,ADMX:2001nej,ADMX:2009iij,ADMX:2018gho,ADMX:2018ogs,ADMX:2019uok,Lee:2020cfj,HAYSTAC:2018rwy,HAYSTAC:2020kwv,Dixit:2020ymh,Alesini:2020vny,Cervantes:2022yzp,DOSUE-RR:2022ise,An:2022hhb,Knirck:2018ojz,Tomita:2020usq,Ramanathan:2022egk} and stellar cooling (`Solar') \cite{Redondo:2013lna,2015JCAP} (additional constraints from tests of Coulomb's law~\cite{williams1971new,bartlett1988limits,tu2004mass,kroff2020constraining} and CROWS~\cite{Betz:2013dza} that are relevant for $\kappa\simeq 10^{-7}$ are not shown).}
    \label{fig:WDDPGCpopulation}
\end{figure*}

In this paper we investigate the conversion of DPs to photons in compact stars. 
We derive, for the first time, the equations governing this process in an anisotropic plasma in the presence of a possibly strong magnetic field. Although not required for conversion, such a magnetic field can have important effects. 
We analyse the resulting signals and detector sensitivity from NSs and also accreting white dwarfs (WDs), which we point out are well suited to conversion. 
As shown in Fig.~\ref{fig:WDDPGCpopulation}, searches for signals from WDs with the upcoming telescopes SKA, GBT and ALMA, operating in GHz to THz frequencies, could surpass current constraints on the kinetic mixing for a wide range of DP  masses. 

The remainder of this paper is structured as follows. In Sec.~\ref{sec:theory} we provide a schematic overview of the resonant conversion process in plasma. In Sec.~\ref{sec:nsconversion} and~\ref{sec:awdconversion} we describe the environments of neutron stars and white dwarfs and details of the conversion process there. In Sec.~\ref{sec:sensitivity} we study the sensitivity of radio telescope to dark photon dark matter. The uncertainties on the dark matter profile are discussed in Sec.~\ref{sec:dmprofile} and the white dwarf environments are revisited in Sec.~\ref{sec:WDenvs}. Finally, in Section~\ref{sec:discussion} we discuss our results and describe future refinements to our analysis. Technical material is provided in the Appendices: we derive the equations governing the conversion in generality and show how these reduce to the expressions in the main text. We also discuss the propagation of photons after production and analyse the impact of additional processes that can affect the conversion. 

\section{Theoretical Framework}
\label{sec:theory}

We consider a DP with Lagrangian density
\begin{equation} \label{eq:L0}
\mathcal{L} = -\frac{1}{4} F'_{\mu\nu}F'^{\mu\nu} + \frac{1}{2} \mAp^2 A'_\mu A'^{\mu} + \frac{\kappa}{2} F'_{\mu\nu}F^{\mu\nu} + \mathcal{L}_{\rm SM}  ~,
\end{equation}
where $F$ ($F'$) is the SM photon (DP) field, and we assume that the dynamics that give rise to the DP mass $\mAp$ are decoupled.    
The kinetic mixing, with coupling $\kappa$, allows conversion between photons and DPs. In a stellar environment this process can be enhanced in the presence of an electron plasma, the properties of which are affected if there is a magnetic field.\footnote{A magnetic field $\gtrsim 4\pi B_c/\alpha$, where $B_c$ is the critical QED magnetic field also leads to non-linear effects through the electron box diagram~\cite{schwinger1951gauge,Fortin:2019npr}, but these are negligible even for NSs.} 
The dynamics of the plasma are described by the permittivity tensor $\pmb{\chi}^p$, reviewed in Appendix~\ref{sec:NSconversion}. 
DPs and photons of energy $\omega$ propagating in the $z$ direction evolve according to
\begin{equation}
    \left[\omega^2+\partial_z^2+\omega^2\begin{pmatrix}
    \pmb{\chi}^p-\mathcal{D}^2&\kappa\pmb{\chi}^p\\
    \kappa\pmb{\chi}^p& -\mAp^2/\omega^2
    \end{pmatrix}\right]
    \begin{pmatrix}
    \pmb{A}\\
    \pmb{A}'
    \end{pmatrix}=0\,,
    \label{eq:combinewaveeq}
\end{equation}
where $\pmb{A}{}^({}'{}^)\equiv(A_x{}^({}'{}^),A_y{}^({}'{}^),A_z{}^({}'{}^))$ 
and $\mathcal{D}^2\pmb{A}\equiv \nabla(\nabla\cdot \pmb{A})/\omega^2$. Conversion is efficient in locations where the photon and DP dispersion relations match, and in these regions the fields can be written as ${\pmb A}{}^({}'{}^)({\pmb x},t)=\tilde{\pmb A}{}^({}'{}^)({\pmb x})e^{i\omega t-ikz}$. To calculate the photon field sourced by DPs we use the WKB approximation. This gives, schematically, 
\begin{equation}
    i\partial_s \tilde{A}_j=\dfrac{1}{2k}(\mAp^2-h_j(s))\tilde{A}_j+\kappa g_j(\tilde{\pmb{A}}')\,,
    \label{eq:delAjgeneral}
\end{equation}
where $s$ is the direction in which the amplitude of the photon field increases, which may not coincide with $z$ in an anisotropic plasma. The relevant photon polarisation, labeled $j$, along with the functions $h_j$ (set by the plasma frequency) and $g_j$ (set by the mixing of DPs and photons) depends on the particular environment. Eq.~\eqref{eq:delAjgeneral} has  solution
\begin{equation}
    i\tilde{A}_j(s)\simeq\int_{0}^sds'g_j(\tilde{\pmb{A}}')e^{i \int_{0}^{s'}ds^{''}(\mAp^2-h(s''))/(2k)}\,
    \label{eq:Ajsolgeneral}\,.
\end{equation}
The integrand in Eq.~\eqref{eq:Ajsolgeneral} is highly oscillatory so the dominant contribution is from the position where the phase in the exponential is stationary $s=s_c$, \ie
\begin{equation}
    \mAp^2=h_j(s_c)\sim \omega_p^2\,,
    \label{eq:resonantcondition}
\end{equation}
which sets the condition for resonant conversion. In what follows we make the simplifying assumption that the photon and DP both travel on exactly radial trajectories in the conversion region, which for an approximately isotropic plasma implies  $\partial_s=\partial_r$, where $r$ is the distance from the star's centre. For an isotropic plasma the effects due to the true trajectories not being exactly radial are small.  Moreover, we expect that the corrections due to the non-isotropic environment of an accreting WD are relatively small although future detailed modeling would be valuable. We assume these relations also hold in NSs, although there may be important effects in this case~\cite{Witte:2021arp,Millar:2021gzs}. 
Additional details of the conversion process are provided in Appendices~\ref{app:neutronstar} and~\ref{app:WDconversion}, and the validity of our assumptions is examined in Appendix~\ref{sec:dephasing}.

\section{Conversion in Neutron Stars}
\label{sec:nsconversion}

We describe NS magnetospheres by the Goldreich--Julian (GJ) model~\cite{Goldreich:1969sb}, which is believed to be accurate in the vicinity of the star~\cite{cohen1972pulsar,hu2021axisymmetric,philippov2015ab}. The charge density at position $\pmb{r}$ above a NS's surface is 
$
    n_{\rm GJ}\simeq|2\pmb{\Omega}\cdot \pmb{B}(\pmb{r})/e|
$,  
where the angular velocity $\pmb{\Omega}$ is related to the spin period $P$ by $|\pmb{\Omega}|=2\pi/P$ (we drop a term of relative importance $\sim |\pmb{\Omega}| r$, which is negligibly small). The magnetic field $\pmb{B}(\pmb{r})$ has a dipolar distribution such that its projection on the rotation axis  $\pmb{\Omega}\cdot \pmb{B}(\pmb{r})/|\pmb{\Omega}| = \frac{1}{2}B_0 \left(r_0/r\right)^3\beta$, where $r_0$ the radius of NS, and $\beta\equiv 3\cos\theta_p\uvec{\pmb{m}}\cdot\uvec{\pmb{r}}-\cos\theta_m$. Here $\theta_p$ is the angle between the position vector $\pmb{r}$ and the rotation axis, $\theta_m$ is the angle between the magnetisation axis $\hat{\pmb{m}}$ and the rotation axis, and $\uvec{\pmb{m}}\cdot \uvec{\pmb{r}}=\cos\theta_m\cos\theta_p+\sin\theta_m\sin\theta_p\cos\Omega t$.
 
We assume the free electron number density $n_e=n_{\rm GJ}$; the resulting plasma frequency is
\begin{equation}
    \omega_p(r)\simeq 2\pi \left(\dfrac{8\pi^3\alpha|\beta|}{em_e}\dfrac{B_0}{P}\right)^{1/2}\left(\dfrac{r_0}{r}\right)^{3/2}\,,
\end{equation}
 For typical NSs $B_0\sim 10^{14}$~G and $P\sim 1$~s, $\omega_p\lesssim 70$~$\mu$eV \ie $\omega_p/(2\pi)\lesssim 17$~GHz.

The typical strong magnetic field in a NS crucially affects dark photon conversion through its effects on the plasma. In the presence of such a field, 
only DP polarisations in the plane spanned by the DP propagation direction and the magnetic field can efficiently convert to photons, and therefore the resultant photon signals are polarised. As before, we take the DP to be propagating in the z-direction, and we fix the magnetic field to lie in the y-z plane at an angle $\theta$ to $\hat{\pmb{k}}$. The induced photon field has a transverse polarisation $A_y$ and a longitudinal polarisation $A_z$ that are interwoven with $A_z=-\mAp^2\cot\theta A_{y}/\omega^2$ (see Appendix~\ref{app:neutronstar} for details).  In the non-relativistic limit $\pmb{A}$ is aligned with $\pmb{B}$ and its amplitude increases in a direction $s$ that is orthogonal to $\pmb{B}$. The photon field's dispersion relation implies a superluminal phase velocity 
and, given its polarisation, it therefore corresponds to the Langmuir-O (LO) mode~\cite{gedalin1998long}, which evolves adiabatically into transverse waves as it propagates out the NS.

The wave equation of $A_y$ is given by Eq.~\eqref{eq:combinewaveeq} with
\begin{equation}
    h_y=\xi\omega_p^2\,,\ g_y=\dfrac{\omega_p^2}{2k}(-\sin^2\theta \tilde{A}'_y+\xi\cot\theta\tilde{A}'_z)\,,
\end{equation}
where $\xi=\sin^2\theta/(1-\frac{\omega_p^2}{\omega^2}\cos^2\theta)$. Hence, the resonance condition is $\mAp^2=\xi\omega_p^2$ and, approximating $\omega\simeq \mAp$, the resonant conversion radius
\begin{equation}
    r_c\simeq r_0\left(\dfrac{8\pi^2\alpha|\beta|}{em_eP}\dfrac{B_0}{\mAp^2}\right)^{1/3}\,.
    \label{eq:rc}
\end{equation}
The conversion probability
\begin{equation}
    p_{\rm NS}\simeq \dfrac{\pi\kappa^2\mAp^3(\cos\theta-\sin^3\theta)^2}{6k|\partial_r\omega_p|\sin^2\theta}\simeq \dfrac{\pi\kappa^2\mAp r_c\beta'}{9v_c}\,,
    \label{eq:pNSratiomain}
\end{equation}
where $\beta'\equiv(\cos\theta-\sin^3\theta)^2/\sin^2\theta$ and $v_c$ is the DP velocity at $r_c$.  
We assume the DP velocity has a  Maxwell-Boltzmann distribution in the galactic rest frame, $f_v(v)\simeq  (\pi v_0^2)^{-3/2} e^{-v^2/v_0^2}$. Starting from an asymptotic velocity $v_i$ far away from NS of mass $M_{\rm NS}$, the infalling DP accelerates to
\begin{equation}
    v_c\simeq \sqrt{\dfrac{2G_NM_{\rm NS}}{r_c}}= 5\times 10^4~{\rm km/s}\sqrt{\dfrac{M_{\rm NS}}{M_\odot}\dfrac{100~\rm km}{r_c}}\,,
\end{equation}
near $r_c$, so $v_c\gg v_i$. By Liouville's theorem, the DP's phase space density is conserved during infalling, so its density near $r_c$ is enhanced to 
$
\rho_{A'}(r_c)=2v_c (\sqrt{\pi} v_i)^{-1}\rho_{A'}^\infty
$,
where $\rho_{A'}^\infty$ is the DP energy density away from NS~\cite{Millar:2021gzs,Leroy:2019ghm}.  

Resonant conversion then yields a photon power per solid angle $\Omega$
\begin{equation}
    \dfrac{d\mathcal{P}}{d\Omega}\simeq 2p_{\rm NS}r_c^2\rho_{A'}(r_c)v_c=\dfrac{4}{\sqrt{\pi}} p_{\rm NS}\rho_{A'}^\infty r_c^2 v_c \dfrac{v_c}{v_i}\,,
    \label{eq:dPdOmegaNS}
\end{equation}
where the factor of 2 accounts for conversion when approaching and leaving the NS. We assume that the magnetisation axis aligns with the rotation axis, \ie~$\theta_m=0$, so in the GJ model $\beta=3\cos^2\theta_p-1$ and  $\cos\theta=2\cos\theta_p/\sqrt{3\cos^2\theta_p+1}$. 
The conversion probability diverges as $\theta$ tends to 0 due to the relation between $A_y$ and $A_z$. This would be regulated by the inclusion of vacuum polarisation effects, the variation of the resonance condition within the conversion length, or the back conversion of photons to DPs. However, we simply impose $p_{\rm NS}(\theta)\leq 1000 p_{\rm NS}(\pi/2)$ or $1.8^\circ\leq\theta\leq 178.2^\circ$, which is expected to be conservative as it leads to $p_{\rm NS}(\theta)\ll 1$ in all the parameter space of interest. 

When considering the signal from a collection of stars, we average over the $\theta_p$ angular dependence in Eq.~\eqref{eq:dPdOmegaNS}.  
We also average over the asymptotic dark photon velocity: $\langle 1/v_i\rangle = \int f_v(v_i)/v_i d^3v_i\simeq 2/(\sqrt{\pi}v_0)$. The mean emission power per NS is then
\begin{equation}
\begin{split}
    &\dfrac{d\mathcal{P_{\rm NS}}}{d\Omega} =1.7\times 10^7~{\rm W} \left(\dfrac{\kappa}{10^{-8}}\right)^2\left(\dfrac{\rho_{A'}^\infty}{0.3~{\rm GeV/cm}^3}\right)\\
    &\left(\dfrac{10^{-5}~\rm eV}{\mAp}\right)^{2/3}\left(\dfrac{300~\rm km/s}{v_0}\right)
    \left(\dfrac{B_0}{10^{14}~\rm G}\dfrac{1~\rm s}{P}\right)^{5/6}\,,
\end{split}
\label{eq:dPdOmegaNSmean}
\end{equation}
where we take $r_0=10$~km and $M_{\rm NS}=M_\odot$. The produced photons travel out of the neutron star with negligible absorption or scattering.

\section{Conversion in Accreting White Dwarfs}
\label{sec:awdconversion}

Mass accretion onto a WD from a companion main sequence star converts gravitational energy to heat and produces a hot and dense plasma~\cite{mukai2017x}. We focus on non-magnetic accreting WDs, in particular 
non-magnetic cataclysmic variables (CVs). In these, the accreting mass forms a disk, which, near the surface of the WD, is decelerated resulting in a boundary layer. 
If the accretion rate $\dot{M}\lesssim 10^{16}~{\rm g/s}$, the boundary layer is thought to be an optically thin plasma that is heated to a temperature $T_s\simeq 10^8$~K, explaining the observation of hard X-rays from such systems~\cite{pringle1979x,king1984x,patterson1985x,frank2002accretion}. The boundary layer extends from the surface of the WD at $r=r_0$ up to $r_0+b$. Throughout this region the gravitational potential is balanced by the radial pressure gradient, which implies~\cite{frank2002accretion}
\begin{equation}
    b\simeq 600~{\rm km}~ \left( \frac{T_s}{10^8~\rm K} \right) \left( \frac{M_{\rm WD}}{M_\odot} \right) \left(\dfrac{r_0}{0.01~R_\odot}\right)^2\,.
\end{equation}
$b$ also sets the scale over which physical properties vary in the radial direction, \ie $\partial/\partial r\sim 1/b$~\cite{frank2002accretion}. 
We assume the $\alpha$-disk model~\cite{shakura1973black,frank2002accretion}. In this,  the disk's scale height at the outer edge of the boundary layer
\begin{equation}
    H=2\times 10^{3}~{\rm km}~\alpha_d^{-1/10}\dot{M}_{16}^{3/20} \left(\frac{r_0+b}{10^{5}\,\rm km}\right)^{9/8}f_r^{3/5}\,,    
\end{equation}
and the matter density at the centre of the disk just outside the boundary layer
\begin{equation}
    n_d=2\times 10^{16}~{\rm cm}^{-3}\alpha_d^{-7/10}\dot{M}_{16}^{11/20}\left(\frac{r_0+b}{10^{5}\,\rm km}\right)^{-15/8}f_r^{11/5}\,,
\end{equation}
where $\dot{M}_{16}=\dot{M}/(10^{16}{\rm g/s})$, $\alpha_d$ parameterises the disk viscosity (we set this to 1), $f_r=(1-\sqrt{r_0/(r_0+b)})^{1/4}$, and we fix $M_{\rm WD}=M_\odot$. 
The matter density in the transverse direction drops as $ n_d \exp(-h^2/H^2)$, where $h$ is the distance perpendicular to the disk. 
Given that the boundary layer is fed by the accretion disk, we assume that the electron density inside the boundary layer has the same transverse profile, \ie  ~\cite{patterson1985x}
\begin{equation}
    n_e=n_d\exp\left(1-\dfrac{r-r_0}{b}-\dfrac{h^2}{H^2}\right)\,,
\end{equation}
and that the  temperature is constant throughout.

Because there is not a strong magnetic field, the longitudinal polarisation of the photon does not propagate in the boundary layer plasma and only conversion of transverse DPs is relevant for the signal. This is described by Eq.~\eqref{eq:delAjgeneral} with
\begin{equation}
    h_x=h_y=\omega_p^2\,,\ g_x=g_y=-\dfrac{\omega_p^2}{2k}\,.
\end{equation}
The resonance condition is $\mAp^2=\omega^2_p$, which sets the conversion radius to be
\begin{equation}
    r_c=r_0+b~\mathrm{ln}\left(\dfrac{4\pi\alpha}{m_e}\dfrac{n_0}{\mAp^2}\right)\simeq r_0\,,
\end{equation}
where $n_0(h)=n_d\exp(1-h^2/H^2)$. The resulting dark photon-photon conversion probability is
\begin{equation}
    p_{\rm WD}\simeq \dfrac{\pi\kappa^2\omega_p^3}{3k\partial_r\omega_p}=\dfrac{2\pi}{3v_c}\kappa^2\mAp b\,.
    \label{eq:pAAinftyWD}
\end{equation}
$n_0$ does not enter $p_{\rm WD}$ and instead simply sets the maximum DP mass for which conversion is possible. Resonant conversion occurs for $\mAp^2m_e/(4\pi\alpha) \leq n_d$, taking place on both sides of the disk. 

The photons produced can be absorbed by inverse bremsstrahlung as they travel out of the WD, and we define $p_s^{\rm{IB}}$ to be the survival probability (an explicit expression is given in Appendix~\ref{sec:attenuation}). Additionally, because the boundary layer is not exactly isotropic, the photons will be deflected slightly in the direction of the density gradient, however we leave a detailed modelling for future work and continue to assume exactly radial trajectories. 
Given the boundary layer's finite transverse depth, photons are only emitted in some directions; the power per solid angle along these is 
\begin{equation}
\begin{split}
&\dfrac{d\mathcal{P}_{\rm WD}}{d\Omega}
    =2.3\times 10^{11}~{\rm W}\ \left(\dfrac{\kappa}{10^{-8}}\right)^2
    \left(\dfrac{\mAp}{10^{-5}~\rm eV}\right)\\
&\left(\dfrac{\rho_{A'}^\infty}{0.3~{\rm GeV/cm}^3}\right)
    \left(\dfrac{300~\rm km/s}{ v_0}\right) \left( \frac{T_s}{10^8~\rm K} \right)  p_s^{\rm{IB}}\,,
    \end{split} \label{eq:PAWD}
\end{equation}
analogously to Eq.~\eqref{eq:dPdOmegaNS}, where we have fixed $r_c\simeq r_0=0.01R_\odot$ and $M_{\rm WD}=M_\odot$. 

\section{Signals and Detection Sensitivity}
\label{sec:sensitivity}
We consider the radio signals from compact stars in the GC,  where the dark matter density is greatly enhanced relative to the Earth's local environment. To quantify the uncertainties from the dark matter density distribution, 
we compare the signals from two representative profiles: the generalised Navarro–Frenk–White (gNFW) profile~\cite{Benito:2016kyp,Benito:2020lgu} and a density spike near the central black hole~\cite{Lacroix:2018zmg,Bertone:2002je,Lacroix:2013qka}. However, we note that a cored profile is not ruled out~\cite{Bullock:2017xww}, and, as we will discuss in Sec.~\ref{sec:dmprofile}, would lead to weaker limits. We assume that the DP makes up the entirety of the dark matter abundance.

The dark matter distribution inferred from the circular velocity profile of luminous stars can be well described by a generalised Navarro–Frenk–White (gNFW) profile~\cite{Benito:2016kyp,Benito:2020lgu}
\begin{equation}
    \rho_{\rm halo}(R)=\rho_0 \left(\dfrac{R_0}{R}\right)^\gamma \left(\dfrac{R_s+R_0}{R_s+R}\right)^{3-\gamma}\,,
\end{equation}
where $R$ is the distance to the galactic centre, the dark matter density local to the Earth $\rho_0=0.47$~GeV/cm$^3$, and the Earth's distance to the galactic centre $R_0=8$~kpc. The scale radius $R_s=20$~kpc and the profile index $\gamma=1.03$ from fits to data (the choice of the parameters is motivated by the fit using `CjX' baryonic morphology in~\cite{Benito:2016kyp}, which is also consistent with the more recent analysis~\cite{Benito:2020lgu}). This yields a dark photon dark matter density of $1.03\times 10^5$~GeV/cm$^3$ at a distance 0.1~pc from the galactic centre. 
However, the dark matter density near the galactic centre supermassive black hole, Sgr A$^*$, is highly uncertain. If the central supermassive black hole grows adiabatically, the dark matter density within a pc of the galactic centre can be enhanced by orders of magnitude, forming a dark matter spike~\cite{Lacroix:2018zmg,Bertone:2002je,Lacroix:2013qka} (although such a spike is not guaranteed to form \cite{Ullio:2001fb} and might not survive to the present day \cite{Merritt:2002vj,Gnedin:2003rj}). For non-annihilating dark matter the spike density is characterised by a power low at distances $R<R_{\rm sp}$
\begin{equation}
    \rho_{\rm spike}(R)=\rho_{\rm halo}(R)\left(\dfrac{R}{R_{\rm sp}}\right)^{-\gamma_{\rm sp}}\,,
\end{equation}
where we take the spike extension $R_{\rm sp}=80$~pc, and $\gamma_{\rm sp}=7/3$, which yields a dark photon density of $6.2\times 10^8$~GeV/cm$^3$ at 0.1~pc. The difference between the gNFW and spike profiles gives a quantitative estimate of the uncertainties on the dark matter density in the galactic centre.

The distribution of WDs and NSs in the GC is detailed in Appendix~\ref{sec:profiles}. Because only a small fraction of WDs are accreting, we consider the signal from an individual star. The analysis of Chandra  in~\cite{zhu2018ultradeep} suggests that there are about 11 hard X-ray point sources within $8''$ (0.3~pc) of the GC, which are likely to be a mixture of magnetic and non-magnetic CVs~\cite{zhu2018ultradeep,Xu:2019sus}.  It is reasonable to assume at least one non-magnetic CV will be aligned such that the radio signal is observable given that the X-ray emissions are expected to be similarly directional, and we conservatively consider the signal from a non-magnetic CV at 0.3~pc. We assume a boundary layer temperature $T_s=4.4\times10^8~{\rm K}$ in a WD with mass $M_{\rm WD}=0.83~M_\odot$ and the expected radius $r_0=0.01~R_\odot$~\cite{yuasa2010white,nauenberg1972analytic}, inspired by the study in~\cite{Xu:2019sus}. We also take $\dot{M}_{16}=1$, which yields $b=3188$~km, $H=107$~km, $n_d=4.6\times 10^{17}$~g/cm$^3$. Note that the boundary layer plasma could be partly relativistic at such high temperature, where the relativistic effect will modify the dielectric tensor of the plasma, which in turn affects the dispersion relation and hence the propagation of photons in the plasma~\cite{swanson2003plasma}. We leave a more dedicated study of such effects in future work.
We estimate the importance of absorption of converted photons by inverse bremsstrahlung as they travel out through the boundary layer by assuming a travelling distance of 500~km (absorption in the cold accretion disk is expected to be less efficient). The resulting attenuation is significant for $\mAp \geq 10^{-4}$~eV, but we stress the true effect depends on the production location and a complete model and a full simulation would be required for a fully reliable analysis.
The signal power ${S}_{\rm sig}$ is the energy flux at Earth divided by the bandwidth $\mathcal{B}$, which we take to be the maximum of signal line-width $\mathcal{B}_{\rm sig}$ and the detection bandwidth of a particular telescope $\mathcal{B}_{\rm det}$.
For a single WD, energy conservation gives
$\mathcal{B}_{\rm sig}=\sqrt{\langle(E_{A'}-\langle E_{A'}\rangle)^2 \rangle}\simeq \sqrt{3/8}\mAp v_0^2$, where $\langle E_{A'}\rangle\simeq \mAp(1+3v_0^2/4)$. 

For NSs we consider the collective signal from all stars that are a distance $R$ between $R_{\rm min}$ and $R_{\rm max}$ from the GC. This leads to
\begin{equation}
\begin{split}
        {S}_{\rm sig}  & = \dfrac{1}{\mathcal{B}d^2}\int_{R_{\min}}^{R_{\rm max}}4\pi R^2 n_{\rm NS}(R) dR\times\\
        &\int_{\log_{10}B_{\min}}^\infty f_B d\log_{10}B_0\int f_P d\log_{10}P \dfrac{d\mathcal{P_{\rm NS}}}{d\Omega}\,,
\end{split}
\label{eq:sigtotNS}
\end{equation}
where $d=8$~kpc is the distance of the Earth from the GC. 
We take the distributions of the magnetic field $f_B$ and spin period  $f_P$ of NSs to be log-normal centred at $ \log_{10}B_0/\rm G=12.65$, $ \log_{10}P/\rm ms=2.7$, with standard deviations $\sigma_{ \log_{10}B_0/\rm G}=0.55$ and  $\sigma_{ \log_{10}P/\rm ms}=0.34$, respectively~\cite{Bai:2021nrs,Faucher-Giguere:2005dxp,Bates:2013uma,Lorimer:2006qs}. The lower limit on the integral over $B_0$, $B_{\rm min}$, is defined by $\omega_p(r_0,B_0=B_{\min})=\mAp$ for $\theta_p=\pi/2$ to facilitate resonant conversion. The population of compact stars in the GC has been studied with Monte Carlo simulations~\cite{freitag2006stellar}. For NSs the population distribution $n_{\rm NS}$ can be fit with a power law that is accurate for $R>R_{\min}=0.025$~pc.  We assume the NSs have a radius of 10~km and an average mass of 1.4~$M_\odot$. The signal from a collection of NSs is broader because the frequencies from the individual sources are Doppler shifted differently due to the motion of the stars, leading to a total width $\mathcal{B}_{\rm sig}\simeq \mAp \sigma_s\sim 10^{-3}\mAp$, where $\sigma_s$ is the stars' velocity dispersion~\cite{Safdi:2018oeu}. We conservatively use the velocity dispersion at $R=0.1$~pc, outside which most of the stars reside. 

To set a limit on, or find evidence for, a DP we require 
the signal power to be larger than the minimum detectable flux density of a radio telescope. This is defined as the fluctuation of the telescope receiver output in a frequency band cumulated over the observation time. Given the narrow bandwidth of the signal, the minimum detectable flux could be orders of magnitude below the continuous background~\cite{condon2016essential}. In~Fig.~\ref{fig:WDDPGCpopulation} we plot the sensitivity reach of SKA~\cite{dewdney2013ska1}, GBT~\cite{GBT} and ALMA~\cite{ALMA} with 100 hours of observation of the GC, which together cover a broad  frequency range from 50~MHz to 950~GHz, as described in Appendix~\ref{sec:radiotelescopes}. 
For the signal from NSs we set $R_{\max}=3$~pc, motivated by the field view of GBT  (at high frequencies $\gtrsim 10$~GHz this may require beams to be combined or a prolonged observation time). 
Additionally, the signal from NSs allows us to set constraints on dark photons in the mass window of 15 to 35~$\mu$eV using the flux density limit data (with background correction) based on observations of the GC with GBT~\cite{Foster:2022fxn} in the Breakthrough Listen (BL) project, which covers a range of 2.9~pc from the GC.

We note that non-accreting WDs can also lead to resonant DP conversion. One possibility is conversion in a WD's atmosphere, which consists of a dense plasma. Due to the relatively low temperatures, the signals from this environment are weaker than those from an accreting WD, 
but future observations might still surpass the cosmological constraint depending on the assumed dark matter density profile. 
Additionally, some isolated WDs might be surrounded by a hot corona, which would lead to strong signals if present in a sizeable fraction of stars. However, as yet there is no compelling evidence of such corona and instead only upper limits on the would-be plasma densities for particular WDs. Details of the signals from these environments and the resulting detection prospects are presented in Appendix~\ref{sec:attenuation}.

\section{Effects of the dark matter density profile} 
\label{sec:dmprofile}

\begin{figure*}[!htb]
    \centering
    \includegraphics[width=0.48\textwidth]{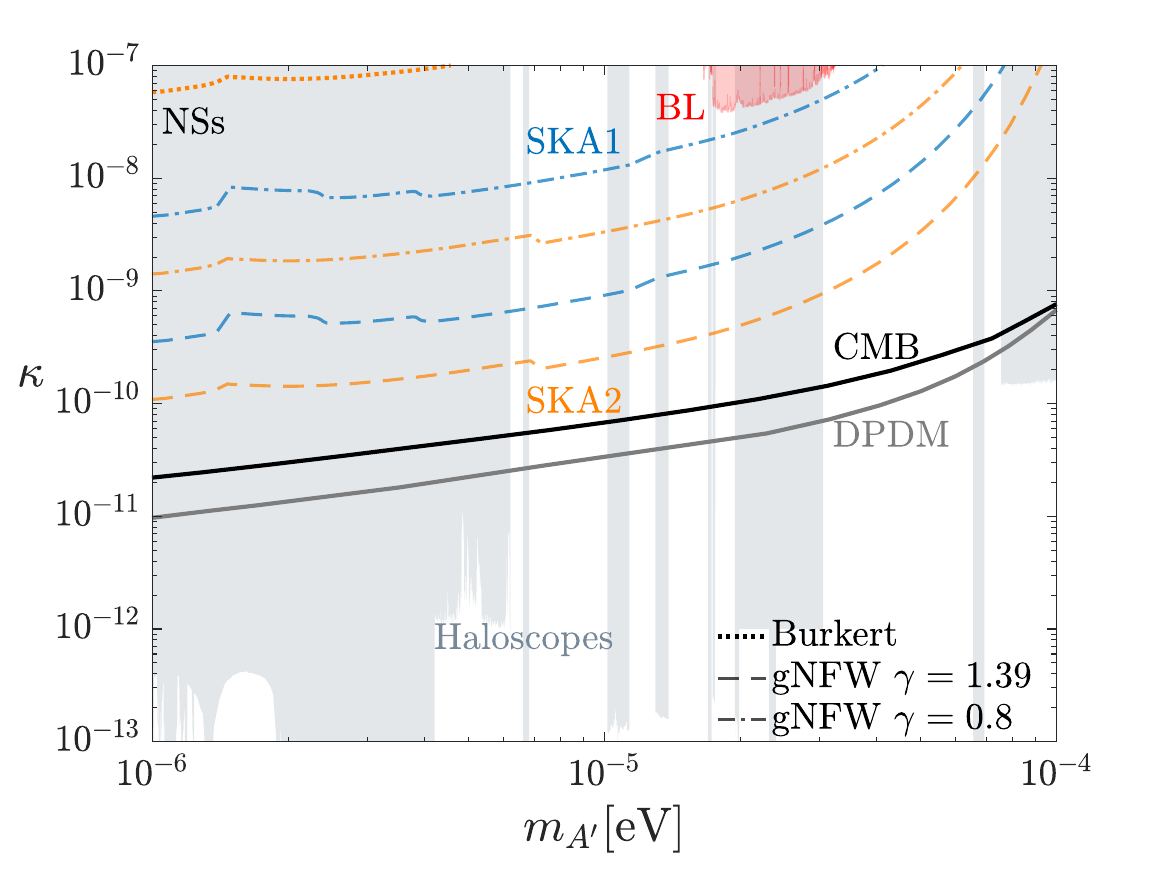}
    \includegraphics[width=0.48\textwidth]{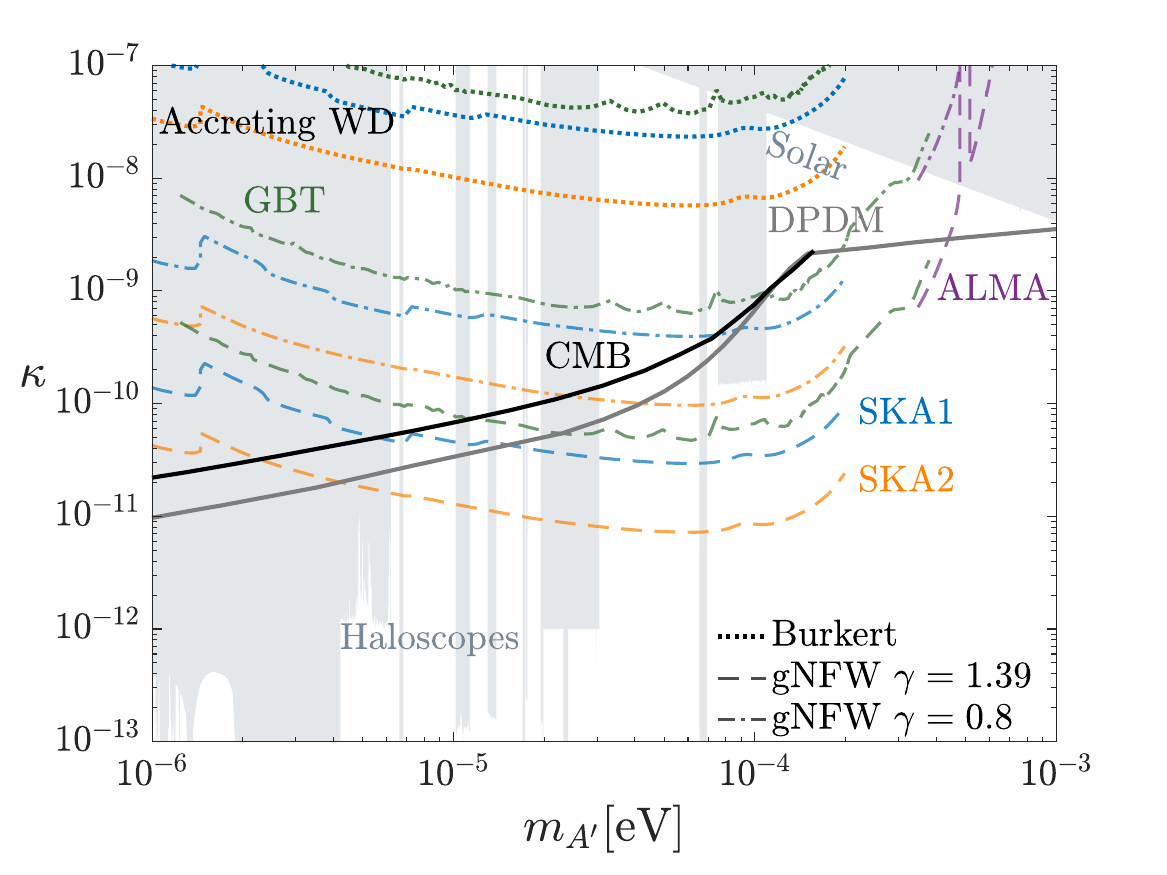} \caption{ The dependence of the sensitivity of radio telescopes to dark photon dark matter on the assumed dark matter spatial distribution. As in Fig.~\ref{fig:NSconversion}, we assume 100 hours of observation of the cumulative signal from neutron stars  within 3~pc of the galactic centre (NSs, left) and the signal from an individual accreting white dwarf  0.3~pc from the galactic centre (WD, right). Coloured lines give the projected sensitivities of ALMA, GBT, SKA1, and SKA2. The red shaded region is our constraint derived from the Breakthrough Listen (BL) project~\cite{Foster:2022fxn}. For each telescope we plot three lines: the dotted lines assume the Burkert dark matter profile, the dashed lines assume the gNFW profile with the `G2 CM' baryonic morphology in~\cite{Karukes:2019jxv} (the BL limit is only visible with this profile), and the dash-dotted lines assume the same profile with the `E2 HG' baryonic morphology in~\cite{Karukes:2019jxv}. Unlike Fig.~\ref{fig:NSconversion}, we do not plot the results assuming a density spike in the galactic center. The grey line depicts the cosmological constraint on dark photon dark matter (DPDM) from Arias~\cite{Arias:2012az}, and the black line shows the constraint from CMB distortion from~\cite{McDermott:2019lch} (CMB).}
    \label{fig:WDDPGCprofiles}
\end{figure*}

In Sec.~\ref{sec:sensitivity} we consider two cuspy dark matter density profiles, the gNFW profile and a density spike in the galactic center. Observation shows galaxies with high stellar density are more likely to be cuspy than cored~\cite{Kaplinghat:2019dhn}, and recent studies varying baryon models indicate that the NFW profile generally fits the rotation curve data better than the cored Burkert profile~\cite{Lin:2019yux} in the Milky Way. Meanwhile, dedicated simulations including baryon feedback suggest that the dark matter profile might be even further contracted close to the galactic center than in an NFW profile~\cite{Cautun:2019eaf} (which, although we do not investigate this possibility in detail, would strengthen our projected sensitivity). However, we note that a cored profile in the Milky Way is not ruled out~\cite{Bullock:2017xww}. To explore the impact of this scenario, we assume the dark matter in the Milky Way follows Burkert profile~\cite{Burkert:1995yz}
\begin{equation}
    \rho_{\rm core}(R)=\rho_s\left(\dfrac{R_c+R}{R_c}\right)^{-1}\left(1+\dfrac{R^2}{R_c^2}\right)^{-1}\,,
\end{equation}
where we take $\rho_s=1.79$~GeV/cm$^3$ and the core radius $R_c=7.8$~kpc, corresponding to the `B4D4C1' baryon model in~\cite{Lin:2019yux} which produces the minimum $\chi^2$ in the fit for Burkert halo. The resulting dark photon sensitivity is displayed in Fig.~\ref{fig:WDDPGCprofiles} with dotted lines. 

We also note that, even assuming a gNFW profile, there is a residual uncertainty on the fit from the choice of baryon model (or morphology). To illustrate the impact of this, in Fig.~\ref{fig:WDDPGCprofiles} we plot the dark photon sensitivity with the alternative gNFW parameters with $\gamma=0.8$ (`E2 HG' morphology in~\cite{Karukes:2019jxv}) and $\gamma=1.39$ (`G2 CM' morphology in~\cite{Karukes:2019jxv}), corresponding to the least and most cuspy dark matter profiles for the baryon models analyzed in~\cite{Karukes:2019jxv}.   

Fig.~\ref{fig:WDDPGCprofiles} show that the least cuspy gNFW profile leads to slightly weaker sensitivity to $\kappa$, while the most cuspy one will enhance the sensitivity of the gNFW profile in Fig.~\ref{fig:NSconversion} by an order of magnitude. The `Breakthough' constraint from neutron stars is already visible even without assuming a density spike in the galactic center in this case. A cored profile, on the other hand, will reduce the sensitivity by about two orders of magnitude compared with the gNFW profile in Sec.~\ref{sec:sensitivity}.

Additionally, although we demonstrate the potential of radio telescopes to discover dark photon dark matter by considering the Milky Way, signals from nearby galaxies are also interesting. It is likely that some nearby galaxies host cuspy dark matter profiles or even density spikes, and these could potentially lead to strong constraints, although we leave an analysis to future work.

\section{Effects of the white dwarf environment} 
\label{sec:WDenvs}

Here we describe our assumptions about the white dwarf environment in more detail and analyse the resulting uncertainties on the projected sensitivity to dark photon dark matter conversion. We focus on non-magnetic cataclysmic variables.

Our assumption that there is at least one accreting white dwarf within $0.3~{\rm pc}$ of the galactic centre is supported by observational evidence. In particular, \cite{zhu2018ultradeep} shows that a significant fraction of the detected hard X-ray point sources in the galactic center is attributable to  the non-magnetic cataclysmic variables (CVs) that we consider, in addition to magnetic CVs (it is also thought that magnetic CVs only make up about 10\% of all CVs~\cite{mukai2017x}). Furthermore, the observed cumulative hard X-ray spectrum can be well fit by thermal bremsstrahlung~\cite{Xu:2019sus}, suggesting that most of the detected X-rays come from the thermal plasmas formed in accretion, with non-magnetic CVs contributing significantly.

\begin{figure*}[!htb]
    \centering
    \includegraphics[width=0.48\textwidth]{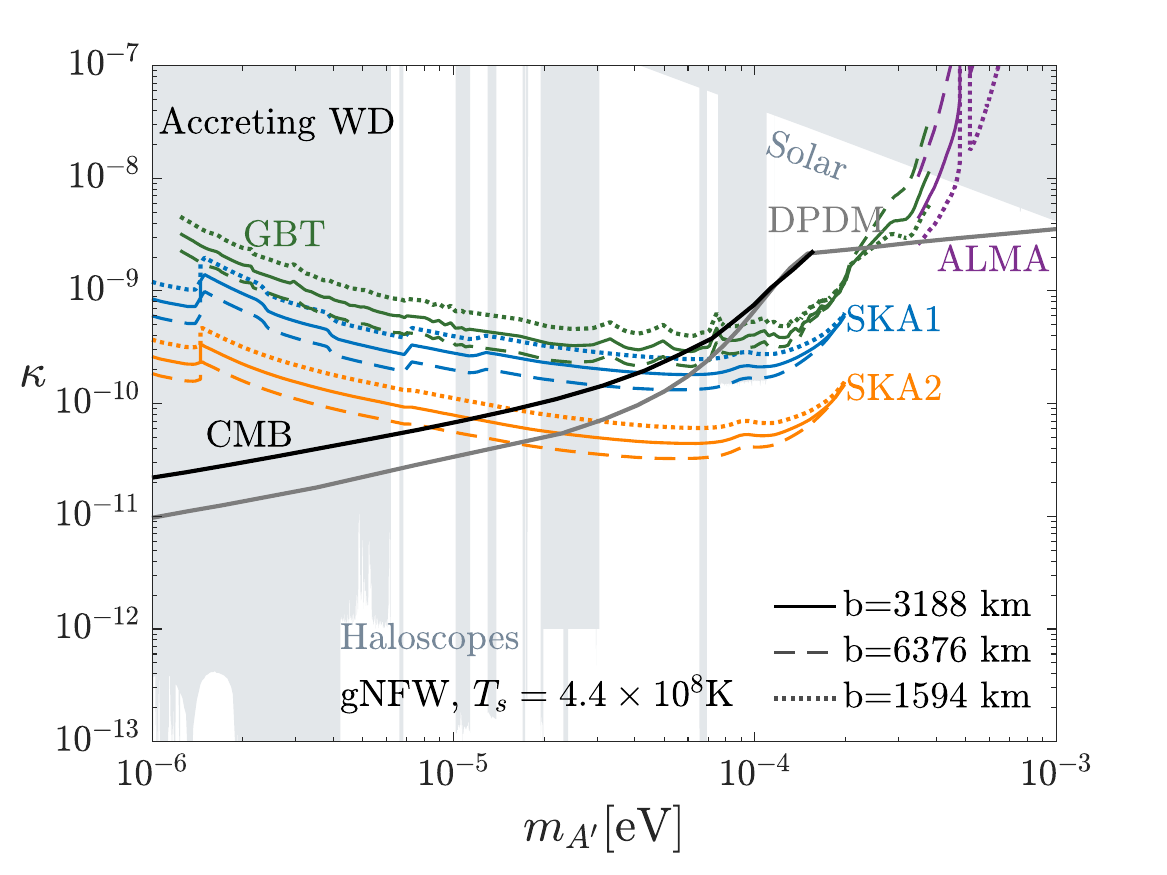}
    \includegraphics[width=0.48\textwidth]{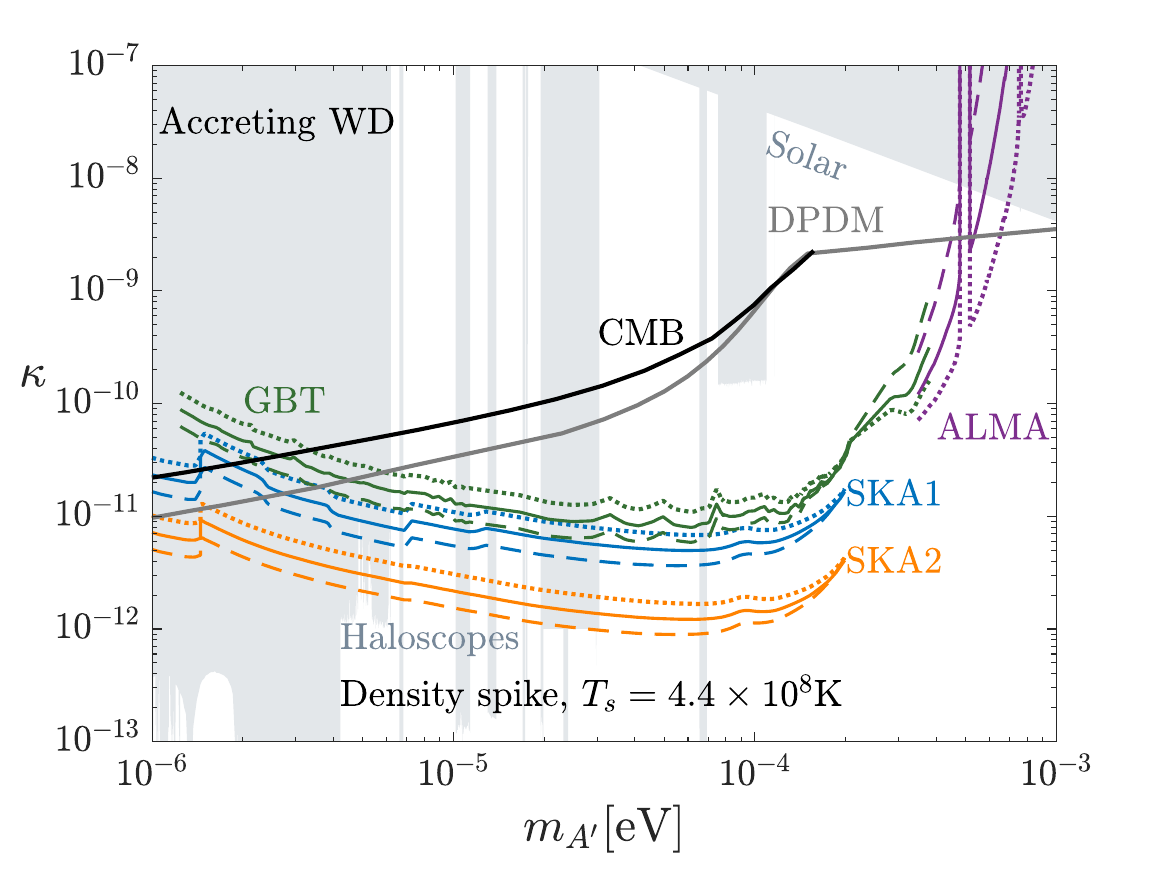} \caption{
The impact of the assumed radial extent of the boundary layer in accreting white dwarfs $b$ on the sensitivity to dark photon dark matter. As in the other figures, we assume 100 hours of observation of the signal from an individual accreting white dwarf  0.3~pc from the galactic centre. The solid, dashed, and dotted lines correspond to the radial extension of the boundary layer $b=3188$~km (adopted in Fig.~\ref{fig:NSconversion}), 6376~km and 1594~km, respectively. The plasma temperature $T_s=4.4\times 10^8$~K is assumed in the boundary layer. {\it Left:} Results obtained assuming the dark matter follows the gNFW profile. {\it Right:} Same as the left, but a density spike is assumed in the galactic center. The grey line depicts the cosmological constraint on dark photon dark matter (DPDM) from~\cite{Arias:2012az}, and the black line shows the constraint from CMB distortion from~\cite{McDermott:2019lch} (CMB). }
    \label{fig:WDDPGCbs}
\end{figure*}

Our analysis also involves assumptions  about the shape of the boundary layer and the electron density distribution. Since dark photons travel approximately in the radial direction, Eq.~\eqref{eq:pAAinftyWD} indicates that the dark photon conversion probability is only related to the derivative of the radial electron density profile, not the electron density in the vertical direction. The scale height $H$ of the boundary layer, inferred from the $\alpha$-disk model, will slightly change the anisotropy of the plasma as well as the emission region, but this has little effect on the resulting signal power. For similar reasons, $n_d$ which describes the matter density only sets the maximum plasma frequency or the maximum dark photon mass, but does not significantly affect the conversion probability. As discussed in Appendix~\ref{app:WDnon}, the radial extension of the boundary layer $b$ is derived from hydrostatic equilibrium where the pressure gradient of the gas balances the gravitational potential, so that $b$ is solely determined by the plasma temperature and the mass of white dwarf. This relation has been used to infer the electron density profile of white dwarf corona~\cite{ingham1976origin,Gill:2011yp,Wang:2021wae} as well as the properties of the boundary layer~\cite{frank2002accretion}.  We stress that the radial electron density profile in the boundary layer is presently uncertain and an important topic for future dedicated study. To estimate the effects of the uncertainty on the density profile, in Fig.~\ref{fig:WDDPGCbs} we plot the sensitivity varying $b$ independently of the plasma temperature. The effects are two-fold. On the one hand, increasing $b$ enhances the conversion probability through the density gradient $\partial_r\omega_p$. On the other hand, a larger $b$ also affects absorption and leads to signal loss at high dark photon mass.

\begin{figure*}[!htb]
    \centering
    \includegraphics[width=0.48\textwidth]{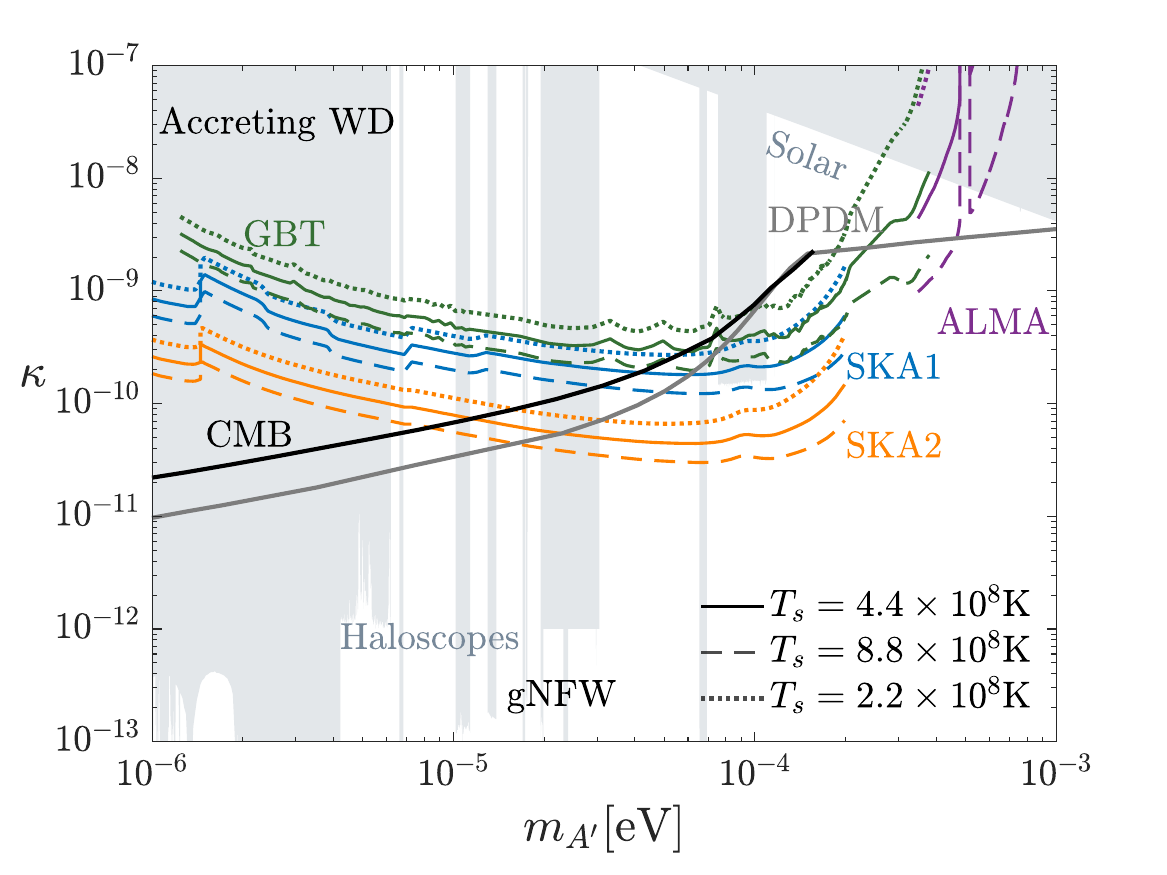}
    \includegraphics[width=0.48\textwidth]{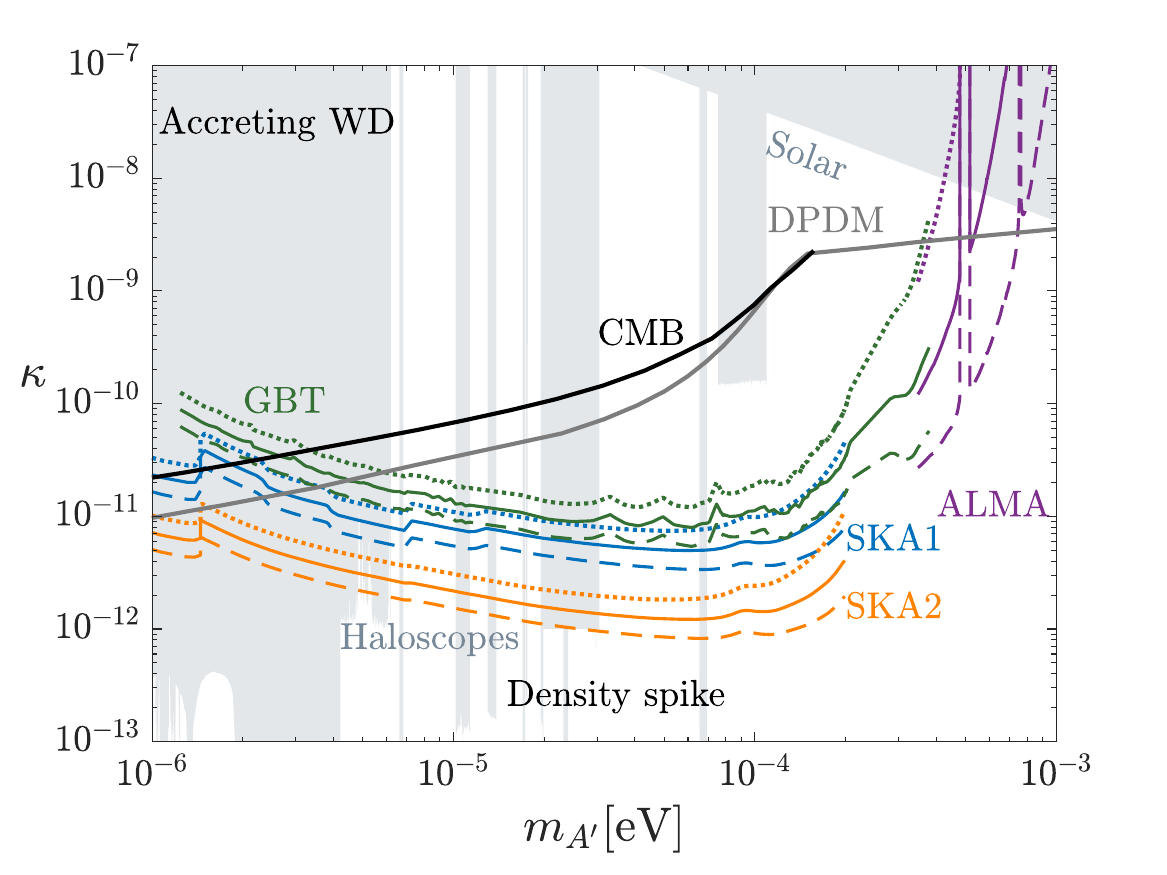} \caption{
The impact of varying the temperature of the accreting white dwarf boundary $T_s$ on the sensitivity of radio telescopes to dark photon dark matter from 100 hours of observation of the signal from an individual accreting white dwarf  0.3~pc from the galactic centre. The solid, dashed, and dotted lines correspond to boundary layer temperatures of $T_s=4.4\times 10^8$~K (adopted in Fig.~\ref{fig:NSconversion}), $8.8\times 10^8$~K and $2.2\times 10^8$~K, respectively. The radial extension of the boundary layer is assumed to be $b=3188$~km. {\it Left:} Results obtained assuming the dark matter follows the gNFW profile. {\it Right:} Same as the left, but a density spike is assumed in the galactic center. The grey line depicts the cosmological constraint on dark photon dark matter (DPDM) from~\cite{Arias:2012az}, and the black line shows the constraint from CMB distortion from~\cite{McDermott:2019lch} (CMB).}
    \label{fig:WDDPGCTs}
\end{figure*}

The temperature of the boundary layer $T_s$ is determined from the flux ratio of the Fe XXVI to Fe XXV emission lines ($I_{7.0}/I_{6.7}$) of the hard X-rays observed in the galactic center~\cite{Xu:2019sus}. We use the low luminosity samples (GCXE-L) which are likely to come from non-magnetic cataclysmic variables instead of the magnetic ones~\cite{Xu:2019sus,zhu2018ultradeep}. The resulting inferred temperatures range from 30 to 50~keV, with a mean of 38~keV ($4.4\times 10^8$~K). To illustrate the effects of a different $T_s$ we allow the temperature to vary by a factor of 2 in Fig.~\ref{fig:WDDPGCTs} (we also restore the relation between $T_s$ and $b$ to highlight the effect of $T_s$ only). In general, a higher plasma temperature both facilitates conversion and suppresses absorption.

\section{Discussion and Outlook}
\label{sec:discussion}

Future telescopes searching for signals from an accreting WD could  
cover a substantial region of viable parameter space with $\mAp\gtrsim 10^{-5}~{\rm eV}$. 
This is the case even with conservative assumptions about the dark matter distribution, and the sensitivity is greatly enhanced if the dark matter profile has a spike. 
The projected reach from signals from an accreting WD surpasses that from NSs due to the dependence of the emission power on the radius of the resonant conversion region as well as the relatively high temperatures and plasma densities in the boundary layer. However, it will be important to study the WD's properties in more detail in the future, \eg modeling the boundary layer and the accretion disk in detail. A component of the observed X-rays from CVs might be generated by magnetic reconnection~\cite{takata2018non} instead of accretion, and the resulting environment may also lead to interesting signals. In addition, since the boundary layer may not be exactly isotropic, signal photons could be refracted when they propagate out of the WD, potentially reducing the signal power. This is to be scrutinized in the future with a more realistic boundary layer profile. The temperature profile of the boundary layer will also affect the absorption of the signals. Finally, we stress that the dark matter distribution in the GC has a strong impact on the projected sensitivity, see Sec.~\ref{sec:sensitivity} and~\ref{sec:dmprofile}, and it will be crucial to improve on this uncertainty in the future.

The signals we have studied complement future haloscope searches, \eg\cite{Gelmini:2020kcu}, which have projected sensitivity to smaller $\kappa$ but can only scan frequencies slowly. The discovery of a radio signal would provide experiments with a DP mass to target while direct detection searches could test the origin of a radio line unaffected by astrophysical uncertainties.  Being independent of dynamics in the early Universe, searches for radio signals are also a useful addition to cosmological constraints, which are also subject to uncertainties and systematics. The more recent analysis of \cite{McDermott:2019lch} gives limits a factor $\simeq 2$ weaker, can be seen from the difference between the `CMB' and `DPDM' lines in Fig.~\ref{fig:WDDPGCprofiles}. In addition, the dark photon signals studied are insensitive to new physics in the early Universe, including the radiation dominated era, whereas the cosmological constraint depends on the dark photon dynamics at redshifts $z\gtrsim 10^5$. 

There are numerous possible extensions to our work. Having set up the formalism in generality, our analysis could be  improved by solving the full 3-dimensional equations and utilising ray-tracing.  The detection sensitivity might be improved by considering globular clusters (which have large concentration of compact stars and low velocity dispersion) \cite{Wang:2021hfb} or nearby galaxies that might have very cuspy dark matter profile. It may also be possible to exploit the fact that the GC NS signal is composed of a forest of ultra-thin lines, or that the signal from a particular NS is polarised~\cite{Dessert:2022yqq}. 
Other possibilities to consider include the signals from collisions of DP substructure such as DP stars~\cite{Gorghetto:2022sue} with astrophysical objects, the changes in theories in which the dynamics that give rise to the DP mass are not decoupled, and whether observable effects occur in theories with different interactions between the DP and the SM.

Finally, there are likely to be interesting signals from axion or DP conversion in other accretion environments. For example, accretion columns form around the magnetic poles in magnetic CVs  and accreting NSs. The densities and temperatures in the resulting plasmas are expected to be similar to those in the boundary layer of non-magnetic CVs, which might allow for interesting signals for axion masses as large as an meV.

\section*{Acknowledgments} 

We thank Richard Battye, Andrea Caputo, Alexander Leder, Tim Linden, Jamie McDonald and Stafford Withington for useful discussions and Marco Gorghetto, John March-Russell and Stephen West for collaboration on related work. We acknowledge the UK Science and Technology Facilities Council for support through the Quantum Sensors for the Hidden Sector collaboration under the grant ST/T006145/1. EH is also supported by UK Research and Innovation Future Leader Fellowship MR/V024566/1. NS is also supported by the National Natural Science Foundation of China (NSFC) Project No. 12047503.

\bibliography{dpradio}

\onecolumngrid
\newpage
\appendix


\section{Dark photon propagation in magnetised plasma}
\label{sec:NSconversion}
Here we derive the equations of motion of photons and dark photons propagating in an anisotropic plasma, such as the magnetosphere of a neutron star, with a (possibly strong) external magnetic field $B$. We address the effects of the plasma and the non-linear interactions induced by a strong magnetic field in turn. Following the conventions of Ref.~\cite{Fortin:2019npr} we write the relevant parts of the photon and dark photon Lagrangian as
\begin{equation}
\mathcal{L} =-\frac{1}{4} (F_{\mu\nu}F^{\mu\nu} +F'_{\mu\nu}F'^{\mu\nu}) + \frac{\kappa}{2} F'_{\mu\nu}F^{\mu\nu}+\frac{1}{2} \mAp^2 A'_\mu A'^{\mu} -A_\mu J^\mu\,.
\end{equation}
It is convenient to redefine the photon field $A_\mu\rightarrow A_\mu+\kappa A'_\mu$ to remove the mixing term, which yields
\begin{equation}
\mathcal{L} =-\frac{1}{4} (F_{\mu\nu}F^{\mu\nu} +F'_{\mu\nu}F'^{\mu\nu})+\frac{1}{2} \mAp^2 A'_\mu A'^{\mu}+\mathcal{L}_a+\mathcal{O}(\kappa^2)\,,
\label{eq:LEM}
\end{equation}
with $\mathcal{L}_a=-\bar{A}_\mu J^\mu$
and $\bar{A}_\mu=A_\mu+\kappa A'_\mu$. The dark photon coupling to electrons is suppressed by $\kappa$ and therefore weak in our parameter space of interest. $\bar{A}_\mu$ is the active state that interacts with the electromagnetic current $J^\mu$. In an anisotropic plasma, the Lagrangian of the current is modified to~\cite{vanderlinde2004classical}
\begin{equation}
    \mathcal{L}_a=-\bar{A}_\mu J^\mu_f+\dfrac{1}{2}\bar{F}_{\mu\nu}\bar{ P}^{\mu\nu}\,, \label{eq:La}
\end{equation}
where $\bar{F}_{\mu\nu}=F_{\mu\nu}+\kappa F'_{\mu\nu}$, $J^\mu_f$ is the free current density, and $\bar{ P}^{\mu\nu}$ is the polarisation tensor induced by the active state $\bar{A}_\mu$,
\begin{equation}
    \bar{P}^{\mu\nu}=\begin{pmatrix}
    0&\bar{P}_x&\bar{P}_y&\bar{P}_z\\
    -\bar{P}_x &0&-\bar{M}_z&\bar{M}_y\\
    -\bar{P}_y &\bar{M}_z&0&-\bar{M}_x\\
    -\bar{P}_z &-\bar{M}_y&\bar{M}_x&0
    \end{pmatrix}\,.
    \label{eq:Pmunu}    
\end{equation}
Because the plasma is not expected to be ferromagnetic, we assume the magnetisation $\bar{\pmb M}=(\bar{M}_x,\bar{M}_y,\bar{M}_z)=0$,  although the collective motion of electrons could potentially induce magnetisation, see \eg 
\cite{shukla2010magnetization}. As in the main text, we assume that both the dark photon and photon propagate in the $z$ direction, and we write their fields in the wave form $\pmb{A}{}^({}'{}^)({\pmb x},t)=\tilde{\pmb A}{}^({}'{}^)({\pmb x})e^{i\omega t-ikz}$. The polarisation induced by the active field is related to the electric fields by the electric permittivity tensor $\pmb{\mathbf{\chi}^p}$
\begin{equation}
    \bar{P}_i=\chi_{ij}^{p}(E_j+\kappa E'_j)\,,
\end{equation}
where $j$ is summed over. In turn, the permittivity tensor is determined by the dielectric tensor $\pmb{\epsilon}$~\cite{potekhin2004electromagnetic,Hook:2018iia,Witte:2021arp}
\begin{equation}
    {\pmb \epsilon}=1+{\pmb \chi}^{p}=R^{yz}_\theta\cdot\begin{pmatrix}
    \varepsilon&ig&0\\
   -ig&\varepsilon&0\\
    0&0&\eta
    \end{pmatrix}\cdot R^{yz}_{-\theta}\,,
    \label{eq:epsilon}    
\end{equation}
where we fix the external magnetic field to lie in the $y-z$ plane at an angle $\theta$ from the propagation $z$ direction, and $R^{yz}_\theta$ is the rotation matrix in the $y-z$ plane. The entries in the dielectric tensor read
\begin{equation}
    \varepsilon=1-\dfrac{\omega_p^2}{\omega^2-\omega_c^2}\,,\ g=\dfrac{\omega_p^2\omega_c}{\omega(\omega^2-\omega_c^2)}\,, \ \eta=1-\dfrac{\omega_p^2}{\omega^2}\,.
    \label{eq:dielectricdef}
\end{equation}
In Eq.~\eqref{eq:dielectricdef}, the plasma frequency $\omega_p=\sqrt{4\pi\alpha n_e/m_e}$ (where $n_e$ is the free electron number density) and the electron cyclotron frequency $\omega_c=\sqrt{\alpha}B/m_e $. The equations of motion of $A_\mu$ and $A'_\mu$ that follow from Eqs.~\eqref{eq:LEM} and \eqref{eq:La} are
\begin{eqnarray}
\partial_\mu F^{\mu\nu}&=&J_f^\nu+\partial_\mu \bar{P}^{\mu\nu}\,,\label{eq:delA}\\
\partial_\mu F'^{\mu\nu}&=&-\mAp^2A'^\nu+\kappa J_f^\nu+\kappa\partial_\mu \bar{P}^{\mu\nu}\,.\label{eq:delAp}
\end{eqnarray}
Because no free current is expected in the plasma $\pmb{J}_f=0$, and $\pmb{E}^({}'{}^)=-\partial_t\pmb{A}^({}'{}^)$, so Eq.~\eqref{eq:delA} can be rewritten as
\begin{equation}
    (\omega^2+\nabla^2)\pmb{A}-\nabla(\nabla\cdot \pmb{A})+\omega^2\pmb{\chi}^{p}\cdot(\pmb{A}+\kappa\pmb{A}')=0\,.
    \label{eq:delAvec}
\end{equation}
As expected, in the absence of a dark photon the time derivative of Eq.~\eqref{eq:delAvec} leads to the usual Maxwell equation of the electric field.
Meanwhile, the propagation equation of dark photon, Eq.~\eqref{eq:delAp}, can be written as
\begin{equation}
    \DAlambert^2 A'^\nu-\partial^\nu\partial_\mu A'^\mu+\mAp^2 A'^\nu=\kappa\partial_\mu \bar{P}^{\mu\nu}\,.
    \label{eq:delApcov}
\end{equation}
Applying $\partial_\nu$ to Eq.~\eqref{eq:delApcov} shows that $\partial_\mu A'^\mu=0$ because $\bar{P}^{\mu\nu}$ is anti-symmetric. Up to linear order in $\kappa$, Eq.~\eqref{eq:delApcov} reduces to
\begin{equation}
    (\omega^2+\nabla^2)\pmb{A}'-\mAp^2\pmb{A}'+\kappa\omega^2\pmb{\chi}^p\cdot\pmb{A}=0\,.
    \label{eq:delApvec}
\end{equation}
Approximating $\nabla^2$ by $\partial_z^2$, \ie neglecting the second derivatives that do not involve the propagation direction, and combining Eq.~\eqref{eq:delAvec} and Eq.~\eqref{eq:delApvec} correspond to the photon and dark photon propagation equation in the main text. 

Next we consider the non-linear effects that are induced by a strong magnetic field, which modify the polarisation tensor discussed above. As we will show below, these are negligible for neutron stars, but they could be important in systems with even stronger magnetic fields, low plasma frequency or without a plasma. In the presence of a strong external field, the propagating fields experience non-linear QED effects due to the electron box diagram that couples them to the $B$ field, known as \emph{vaccum polarisation}~\cite{Fortin:2019npr,schwinger1951gauge,adler1971photon,Heyl:1997hr}. This adds a non-linear contribution $\mathcal{L}_{\rm nl}$ to the Lagrangian in Eq.~\eqref{eq:LEM}, which is a function of $I\equiv\bar{F}_{\mu\nu}\bar{F}^{\mu\nu}$ (see~\cite{Heyl:1997hr,Fortin:2019npr} for an explicit expression). 

With the inclusion of $\mathcal{L}_{\rm nl}$, the dielectric tensor and the magnetic permeability tensor can be calculated from
\begin{equation}
    \epsilon_{ij}=\dfrac{\partial^2\mathcal{L} }{\partial E_i\partial E_j}\,,\ \mu_{ij}=-\dfrac{\partial^2\mathcal{L} }{\partial B_i\partial B_j}\,,
\end{equation}
\cite{Heyl:1997hr}. The result is that the vacuum contribution modifies the dielectric tensor in Eq.~\eqref{eq:epsilon} to ${\pmb \epsilon}=1+{\pmb \chi}^{p}+{\pmb \chi}^{\rm vac}$ where
\begin{equation}
    {\pmb \chi}^{\rm vac}=R^{yz}_\theta\cdot\begin{pmatrix}
    a&0&0\\
   0&a&0\\
    0&0&a+q
    \end{pmatrix}\cdot R^{yz}_{-\theta}\,,
    \label{eq:epsilonvac}     
\end{equation}
\cite{potekhin2004electromagnetic}. Additionally, the vacuum contribution induces a magnetisation $\bar{\pmb{M}}$ in Eq.~\eqref{eq:Pmunu} that is related to the magnetic field of photon and dark photon through the permeability tensor, \ie $\bar{M}_i=(1-\mu_{ij})(B_j+\kappa B'_j)$, where
\begin{equation}
    {\pmb \mu}^{-1}=1+R^{yz}_\theta\cdot\begin{pmatrix}
    a&0&0\\
   0&a&0\\
    0&0&a+m
    \end{pmatrix}\cdot R^{yz}_{-\theta}\,.
    \label{eq:muvac}     
\end{equation}
Normalising the magnetic field by the critical QED field strength: $\uvec{b}=B/B_c$ with $B_c=\frac{m_e^2c^3}{e\hbar}=4.414\times 10^{13}$~G, the functions $a$, $q$, and $m$ in Eqs.~\eqref{eq:epsilonvac} and \eqref{eq:muvac} can be fit by
\begin{align}
    &a\simeq -\dfrac{2\alpha}{9\pi}\ln\left(1+\dfrac{\uvec{b}^2}{5}\dfrac{1+0.25487\uvec{b}^{3/4}}{1+0.75\uvec{b}^{5/4}}\right)\,,\label{eq:adef}\\
    &q\simeq\dfrac{7\alpha}{45\pi}\dfrac{\uvec{b}^2(1+1.2\uvec{b})}{1+1.33\uvec{b}+0.56\uvec{b}^2}\,,\label{eq:qdef}\\    
    &m\simeq-\dfrac{4\alpha}{45\pi}\dfrac{\uvec{b}^2}{1+0.72\uvec{b}^{5/4}+0.27\uvec{b}^2}\,,\label{eq:mdef}
\end{align}
which are accurate in both the weak field and $b\gg 1$ limits~\cite{potekhin2004electromagnetic}. 
Near the dark photon-photon conversion region both photons and dark photons are non-relativistic with $k\ll \omega$. Because the magnetic field involves the spatial derivative of $\pmb{A}$ and $\pmb{A}'$, the in-medium magnetisation is suppressed by a factor $k/\omega$ compared with the polarisation density $\bar{\pmb{P}}$. Consequently we set $\pmb{\mu}^{-1}=1$ and leave a full exploration to future work. 

The resulting equations of motion are Eq.~\eqref{eq:delAvec} and Eq.~\eqref{eq:delApvec} with the replacement ${\pmb \chi}^{p}\rightarrow {\pmb \chi}^{p}+{\pmb \chi}^{\rm vac}$ where $\pmb{\chi}^p$ is as defined in Eq.~\eqref{eq:epsilon}. We follow the prescription in~\cite{Millar:2021gzs} and neglect second order derivatives that do not involve $z$ because the plasmas we consider are slowly varying. The wave equations of the photon are 
\begin{align}
    &(\omega^2+\partial_z^2)A_x-\partial_x\partial_z A_z+\omega^2[\xi_x\bar{A}_x+ig\cos\theta\bar{A}_y+ig\sin\theta\bar{A}_z]=0\,,\label{eq:delAx}\\
    &(\omega^2+\partial_z^2)A_y-\partial_y\partial_z A_z+\omega^2[-ig\cos\theta\bar{A}_x+\xi_y\bar{A}_y-\xi_{yz}\bar{A}_z]=0\,,\label{eq:delAy}\\
    &\omega^2A_z-\partial_x\partial_z A_x-\partial_y\partial_z A_y+\omega^2[-ig\sin\theta\bar{A}_x-\xi_{yz}\bar{A}_y+\xi_z\bar{A}_z]=0\,,\label{eq:delAz}
\end{align}
where $\varepsilon'=\epsilon-1$, $\eta'=\eta-1$, with $\epsilon$ and $\eta$ given in Eq.~\eqref{eq:dielectricdef}, and we have defined
\begin{align}
    &\xi_x\equiv\varepsilon'+a\,,\\
    &\xi_y\equiv\varepsilon'\cos^2\theta+\eta'\sin^2\theta+a+q\sin^2\theta\,,\\
    &\xi_z\equiv\varepsilon'\sin^2\theta+\eta'\cos^2\theta+a+q\cos^2\theta\,,\\
    &\xi_{yz}=(\eta'-\varepsilon'+q)\cos\theta\sin\theta\,.
\end{align}
The corresponding wave equations for the dark photon are
\begin{align}
    &(\omega^2+\partial_z^2-\mAp^2)A'_x+\kappa\omega^2[\xi_x A_x+ig\cos\theta A_y+ig\sin\theta A_z]=0\,,\label{eq:delApx}\\
    &(\omega^2+\partial_z^2-\mAp^2)A'_y+\kappa\omega^2[-ig\cos\theta A_x+\xi_y A_y-\xi_{yz}A_z]=0\,,\label{eq:delApy}\\
    &(\omega^2+\partial_z^2-\mAp^2)A'_z+\kappa\omega^2[-ig\sin\theta A_x-\xi_{yz} A_y+\xi_z A_z]=0\,.\label{eq:delApz}
\end{align}
\cref{eq:delAx,eq:delAy,eq:delAz,eq:delApx,eq:delApy,eq:delApz} are analogous to Eq.~(3.9) in~\cite{Fortin:2019npr}, but are valid for $k\ll\omega$ instead of the weak dispersion limit $\omega\simeq k$. Moreover, one can straightforwardly obtain results for any $k/\omega$ by including the magnetisation from the vacuum polarisation.

\section{Dark photon conversion in neutron star magnetosphere} \label{app:neutronstar}

Here we analyse the conversion of dark photons to photons in a typical neutron star environment. 
The magnetic field at the surface of neutron stars ranges from $10^8$~G to $10^{15}$~G~\cite{Reisenegger:2003pj}. This sets the electron cyclotron frequency $\omega_c\sim 0.3-3\times 10^6$~eV, which is much larger than the dark photon masses and plasma frequencies relevant in our present work.  In this limit ($\omega_c\gg \omega,\,\omega_p$) we have $\varepsilon'\simeq 0$, $g\simeq 0$,  and the wave equations of photons~\cref{eq:delAx,eq:delAy,eq:delAz} simplify greatly to
\begin{align}
    &(\omega^2+\partial_z^2)A_x-\partial_x\partial_z A_z+\omega^2a\bar{A}_x=0\,,\label{eq:delAxlargeB}\\
    &(\omega^2+\partial_z^2)A_y-\partial_y\partial_z A_z+\omega^2[(\eta'\sin^2\theta+a+q\sin\theta^2)\bar{A}_y-(\eta'+q)\cos\theta\sin\theta\bar{A}_z]=0\,,\label{eq:delAylargeB}\\
    &\omega^2A_z-\partial_x\partial_z A_x-\partial_y\partial_z A_y+\omega^2[-(\eta'+q)\cos\theta\sin\theta\bar{A}_y+(\eta'\cos^2\theta+a+q\cos^2\theta)\bar{A}_z]=0\,.
    \label{eq:delAzlargeB}
\end{align}
It is straightforward to see that $A_x$ only couples to $A_z$ though the derivative term and hence evolves nearly separately from $A_y$ and $A_z$. In principle, $A_x$ can be induced by dark photon conversion, mediated by the vacuum contribution proportional to $\kappa a$ in Eq.~\eqref{eq:delAxlargeB}. However, it has a dispersion relation $\omega=k$, which is identified as the magnetosonic-t mode. As a result, this mode will never match the energy momentum relation of a massive dark photon and cannot be produced on resonance, so we subsequently set $A_x=0$. 

\begin{figure}
    \centering
    \includegraphics[width=0.5\textwidth]{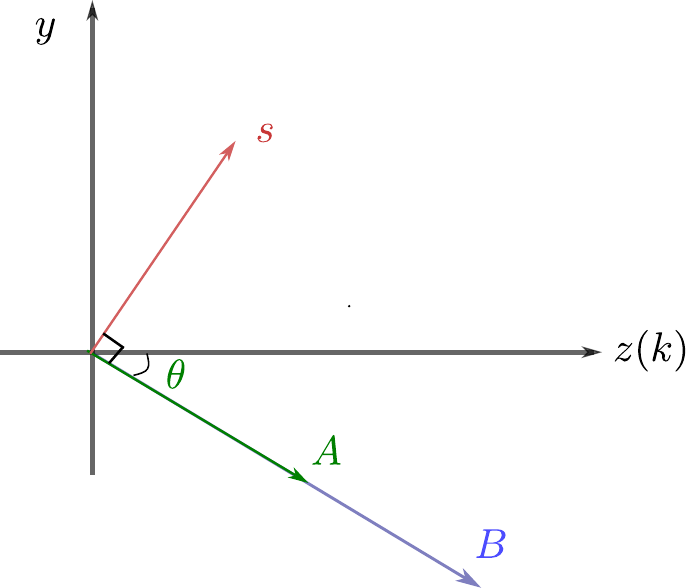}
    \caption{Schematic illustration of dark photon conversion in a magnetised plasma. Dark photons propagate in the $z$ direction, at an angle $\theta$ to the magnetic field. The induced photon field aligns with the magnetic field and its amplitude increases in the $s$ direction, which is perpendicular to the magnetic field.}
    \label{fig:NSconversion}
\end{figure}

Next we consider the induced $A_y$ and $A_z$. First we note that near the conversion region $\omega\sim \omega_p$, $|\eta'|\sim 1$. Therefore, due to the prefactors in Eq.~\eqref{eq:adef} and~\eqref{eq:qdef}, $a\,,q\ll 1$ unless the external magnetic field $B\gtrsim 10^{16}$~G, which is stronger than the maximum expected magnetic field in a neutron star. As a consequence, we neglect all vacuum contributions. To solve the wave equations of photons, we eliminate $A_z$ from Eq.~\eqref{eq:delAylargeB} using Eq.~\eqref{eq:delAzlargeB} to arrive at the differential equation for $A_y$,
\begin{equation}
    \partial_z^2 A_y+\dfrac{2\omega_p^2\sin\theta\cos\theta}{\omega^2-\omega_p^2\cos^2\theta}\partial_y\partial_zA_y+\dfrac{\omega^2(\omega^2-\omega_p^2)}{\omega^2-\omega_p^2\cos^2\theta}A_y=\kappa\left(\omega_p^2\sin^2\theta A'_y-\dfrac{2\omega^2\omega_p^2\sin\theta\cos\theta}{\omega^2-\omega_p^2\cos^2\theta}A'_z\right)\,.
    \label{eq:Aycomb}
\end{equation}
In deriving Eq.~\eqref{eq:Aycomb} from~\cref{eq:delAylargeB,eq:delAzlargeB}, 
we have assumed the plasma frequency varies slowly in the conversion region and that the photon trajectory does not deviate significantly from a straight line due to refraction or gravitational bending, so that the derivatives of $\omega_p$ and $\theta$ can be dropped. We also neglect the derivative of the dark photon field, which is expected to be subdominant during conversion. We can again write the photon and dark photon fields in the wave form $A_i=e^{i\omega t-ikz}\tilde{A}_i(y,z)$, $A'_i=e^{i\omega t-ikz}\tilde{A}'_i(y,z)$ where $\omega^2=k^2+\mAp^2$. Using the WKB approximation with the assumptions $|\partial_z^2\tilde{A}_y|\,,|\partial_y\partial_z\tilde{A}_y|\ll k|\partial_z\tilde{A}_y|\,,k|\partial_y\tilde{A}_y|$, we obtain the first order differential equation
\begin{equation}
    i\partial_s \tilde{A}_y=\dfrac{1}{2k}(\mAp^2-\xi\omega_p^2)\tilde{A}_y+\dfrac{\kappa\omega_p^2}{2k}(-\sin^2\theta \tilde{A}'_y+\xi\cot\theta\tilde{A}'_z)\,,
    \label{eq:delAyds}
\end{equation}
where
\begin{equation}
    \xi=\dfrac{\sin^2\theta}{1-\frac{\omega_p^2}{\omega^2}\cos^2\theta}\,,\ \partial_s=\partial_z+\xi\dfrac{\omega_p^2}{\omega^2}\cot\theta \partial_y\,.
    \label{eq:xiandds}
\end{equation}
This is analogous to the axion-photon conversion described in~\cite{Millar:2021gzs}, except that the source terms are now proportional to $\kappa$ instead of $g_{a\gamma\gamma} B$. As illustrated in Fig.~\ref{fig:NSconversion}, in the non-relativistic limit the converted photons acquire both a transverse component $A_y$ and a longitudinal component $A_z$. 
In combination, the photon polarisation lines up with the direction of the external magnetic field, and evolves in the $s$ direction that is perpendicular to the magnetic field. As mentioned in the main text, the photon mode is identified as the Langmuir-O (LO) mode~\cite{gedalin1998long}. The dispersion relation of the LO mode ($k^2= \omega^2(\omega^2-\omega_p^2)/\left(\omega^2-\omega_p^2\cos^2\theta\right)$) transforms into the free-space dispersion relation as the photon travels outside the magnetosphere and the photon becomes purely transverse. The solution of Eq.~\eqref{eq:delAyds} is
\begin{equation}
    i\tilde{A}_y(s)=e^{if(s)}\int_0^s ds' \dfrac{\kappa\omega_p^2}{2k}(-\sin^2\theta \tilde{A}'_y+\xi\cot\theta\tilde{A}'_z) e^{if(s')}\,,
    \label{eq:Ayintegral}
\end{equation}
where the phase in the exponent
\begin{equation}
    f(s')= \int_{0}^{s'}ds^{''}\dfrac{\mAp^2-\xi\omega_p(s'')^2}{2k}\,.
    \label{eq:exponentialphase}
\end{equation}
The first exponential in Eq.~\eqref{eq:Ayintegral} is a pure phase that does not contribute to the conversion probability. The integrand in Eq.~\eqref{eq:Ayintegral} is highly oscillating and tends to cancel unless the phase is stationary, \ie~$\partial_s f=0$. This gives the resonant conversion condition \begin{equation}
    \mAp^2=\xi\omega_p(s_c)^2\,,
\end{equation}
or
\begin{equation}
    \omega_p^2=\dfrac{\mAp^2\omega^2}{\mAp^2\cos^2\theta+\omega^2\sin^2\theta}\,.
    \label{eq:omegapres}
\end{equation}
The conversion peaks near $s_c$ and for practical purposes can be taken to be vanishing everywhere else. We can therefore expand Eq.~\eqref{eq:exponentialphase} as a Taylor series up to the second order, which yields
\begin{equation}
    f(s')\simeq \int_{0}^{s_c}\dfrac{\mAp^2-\xi\omega_p(s'')^2}{2k}ds''+\dfrac{\pi}{2}\dfrac{(s'-s_c)^2}{L^2}\,,
\end{equation}
where the conversion length is defined as
\begin{equation}
    L\equiv \left(\dfrac{1}{2\pi k}\left\lvert\partial_s(\xi\omega_p(s')^2)\right\rvert\right)^{-1/2}_{s'=s_c}\simeq\dfrac{\sin\theta}{\xi}\sqrt{\dfrac{\pi}{\omega_p\partial_s\omega_p}}\,,
    \label{eq:conversions}
\end{equation}
 We have also neglected the derivative of the dark photon momentum, which only varies slowly due to the gravitational potential of a star. Under this approximation, Eq.~\eqref{eq:Ayintegral} evaluates to~\cite{Millar:2021gzs}
\begin{equation}
    i\tilde{A}_y(s)\simeq \dfrac{\kappa\omega_p^2}{2k}(-\sin^2\theta \tilde{A}'_y+\xi\cot\theta\tilde{A}'_z)\sqrt{2}L\cdot{\rm Erf}\left(\sqrt{-i\frac{\pi}{2}}\frac{s}{2L}\right)\,.
    \label{eq:Aysolution}
\end{equation}
$s$ should be defined at the location beyond which the momenta of photons and dark photons do not match anymore, or where the WKB approximation fails. For $s\gtrsim 2L$, the error function approximately evaluates to 1. For $s\lesssim 2L$, $\sqrt{2}L$ and the error function in Eq.~\eqref{eq:Aysolution} is to be replaced by $\Delta z$, the displacement in $z$ direction where the WKB approximation applies. We refer readers to Refs.~\cite{Millar:2021gzs,Marsh:2021ajy} for more extensive discussions and leave the exploration of this scenario in future work. In the former case, the photon field after conversion
\begin{equation}
    \tilde{A}_y(s)\simeq \dfrac{\kappa}{\mAp^2}\sqrt{\dfrac{\pi\omega_p^3}{2k|\partial_s\omega_p|}}(-\omega_p^2\sin^3\theta \tilde{A}'_y+\mAp^2\cos\theta \tilde{A}'_z)\,.
    \label{eq:Ays}
\end{equation}
Similarly to axions, there are various effects that could modify Eq.~\eqref{eq:Ays}, \eg due to bending of the photon path within the conversion length caused by refraction, gravitational curvature or the variation of the magnetic field. Moreover, if $\partial_s\omega_p$ is close to 0 there is a divergence that will be cut-off by some additional dynamics. We also refer readers to Refs~\cite{Witte:2021arp,Millar:2021gzs} for detailed discussions. From Eq.~\eqref{eq:delAzlargeB}, ignoring the partial derivative and the dark photon mixing terms, we obtain
\begin{equation}
    \tilde{A}_z(s)\simeq -\xi\dfrac{\omega_p^2}{\omega^2}\cot\theta  \tilde{A}_y(s)=-\dfrac{\mAp^2}{\omega^2}\cot\theta \tilde{A}_y(s)\,,
    \label{eq:AztoAy}
\end{equation}
where we have used the exact resonant conversion condition in Eq.~\eqref{eq:omegapres}. The conversion probability
\begin{equation}
    p_{\rm NS}\simeq \dfrac{|\tilde{A}_y|^2+|\tilde{A}_z|^2}{|\tilde{A}'_x|^2+|\tilde{A}'_y|^2+|\tilde{A}'_z|^2}\simeq \dfrac{\pi\kappa^2\omega_p^3(\mAp^2\cos\theta-\omega_p^2\sin^3\theta)^2}{6k\mAp^4|\partial_s\omega_p|\sin^2\theta}\,,
    \label{eq:pNSratio}
\end{equation}
where we have assumed $\mAp\gg k$ and $\tilde{A}'_x\simeq\tilde{A}'_y\simeq \tilde{A}'_z$. The plasma frequency $\omega_p$ at $s_c$ is given by Eq.~\eqref{eq:omegapres}. 

We note that the factor of 
$\cot\theta$ that relates $A_z$ to $A_y$ in Eq.~\eqref{eq:AztoAy} when the full resonant condition Eq.~\eqref{eq:omegapres} is imposed leads to a divergence in $p_{\rm NS}$ in Eq.~\eqref{eq:pNSratio} as $\theta\to 0$. 
This is not dependent on the definition of $L$ because $A_y$ is finite at $\theta=0$, but instead arises from the properties of the plasma which causes the mixing of $A_y$ and $A_z$ in Eq.~\eqref{eq:delAzlargeB}. This divergence would have been artificially removed if we 
we had approximated $\omega=\omega_p=m_a$ (as done in \eg Ref.~\cite{Millar:2021gzs}), which would lead to the relation
\begin{equation}
    \dfrac{E_z}{E_y}=\dfrac{m_a^2\cos\theta\sin\theta}{k^2+m_a^2\sin^2\theta}\,.
\end{equation}
However, there are several caveats to this analysis: 1) We have neglected the vacuum contribution in Eq.~\eqref{eq:delAzlargeB}, the inclusion of which will modify the $\sin\theta$ in the denominator of Eq.~\eqref{eq:AztoAy} and~\eqref{eq:pNSratio}. 2)  Eq.~\eqref{eq:omegapres} varies slightly within the conversion length so the  divergence only appears at $s_c$ and does not hold through the whole conversion region. As $\theta\to 0$ the conversion length $L\to 0$
as well, and Eq.~\eqref{eq:omegapres} holds at this point, so it is not automatic that this regulates the divergence. 3) The conversion probability obviously cannot exceed 1, and when $p_{\rm NS}\sim 1$ photons will convert back to dark photons. 
A dedicated study is required to determine the impact of each of these and to determine the dominant effect that cuts off the divergence. Instead, as mentioned in the main text, we simply require the conversion probability at an arbitrary $\theta$ not to be larger than 1000 times the conversion probability at $\theta=\pi/2$ where only $A_y$ is produced and $A_z$ vanishes. This factor is rather artificial, but, given that we work in the parameter space where $p_{\rm NS}\ll 1$ and $a,q\ll \eta'$, it is reasonable to expect that it is conservative. In the non-relativistic limit, our approximation amounts to imposing $\cot^2\theta\leq 1000$, which yields $1.8^\circ\leq\theta\leq 178.2^\circ$. We leave the exploration of smaller angles to future study. 

Away from the divergence we can safely use the approximation $\omega\simeq \mAp$ to simplify Eq.~\eqref{eq:pNSratio} and obtain
\begin{equation}
    p_{\rm NS}\simeq \dfrac{\pi\kappa^2\mAp^3(\cos\theta-\sin^3\theta)^2}{6k|\partial_s\omega_p|\sin^2\theta}\,. \label{eq:pNSapp}
\end{equation}
Dedicated simulations of their trajectories of dark photons and photons would be required to properly evaluate the conversion probability in Eq.~\eqref{eq:pNSapp}~\cite{Witte:2021arp,Millar:2021gzs}. The approximations we make in the main text of assuming that  both dark photons and photons travel on radial trajectories and that $\partial_s=\partial_r$, amount to effectively neglecting the derivative $\partial_y$ in Eq.~\eqref{eq:xiandds}.  We note that important corrections are likely to arise for specific angles $\theta$ in a proper derivation, as analysed in~\cite{Millar:2021gzs}.

\section{Dark photon conversion in white dwarfs}
\label{app:WDconversion}
Here we consider conversion in white dwarf environments that are either unmagnetised or weakly magnetised (as is the case for the boundary layer in the non-magnetic cataclysmic variables considered in the main text). In this case the opposite limit $\omega_c\ll \omega,\,\omega_p$ to the neutron star analysis applies, which implies $\epsilon'\simeq \eta'=-\omega_p^2/\omega^2$, $g\simeq 0$. \cref{eq:delAx,eq:delAy,eq:delAz,eq:delApx,eq:delApy,eq:delApz} still apply but simplify to
\begin{align}
    &(\omega^2+\partial_z^2)A_i-\partial_i\partial_z A_z-\omega_p^2\bar{A}_i=0\,,\\
    &\omega^2A_z-\partial_x\partial_z A_x-\partial_y\partial_z A_y-\omega^2\bar{A}_z=0\,\\    
    &(\omega^2+\partial_z^2-\mAp^2)A'_j-\kappa\omega_p^2A_j=0\,,
\end{align}
where $i=x,y$, and $j=x,y,z$, and we have set $a=q=0$ because the vacuum contributions are sub-dominant (these equations could also be obtained directly from Eqs.~\eqref{eq:delAvec} and \eqref{eq:delApvec}, neglecting the magnetic field from the start). Therefore, the longitudinal mode of the photon $A_z$ evolves separately and only couples to the transverse modes through the derivative term. The dispersion relation $\omega=\omega_p$ of $A_z$ indicates that it does not propagate, so we drop this in the following analysis. The mixing of the transverse modes can be written in the symmetric form
\begin{equation}
    \left[\omega^2+\partial_z^2+\begin{pmatrix}
    -\omega_p^2&-\kappa\omega_p^2\\
    -\kappa\omega_p^2& -\mAp^2
    \end{pmatrix}\right]
    \begin{pmatrix}
    A_i\\
    A'_i
    \end{pmatrix}=0\,,
\end{equation}
where $i=x,y$. The components $A_x$ and $A_y$ sourced from the corresponding dark photon modes also propagate separately from each other. Using the WKB approximation, we obtain
\begin{equation}
    i\partial_z \tilde{A}_i=\dfrac{1}{2k}(\mAp^2-\omega_p^2)\tilde{A}_i-\dfrac{\kappa\omega_p^2}{2k} \tilde{A}'_i\,,
\end{equation}
which has solution
\begin{equation}
    i\tilde{A}_i(z)\simeq -\kappa\omega_p\sqrt{\dfrac{\pi\omega_p}{2k\partial_z\omega_p}}\tilde{A}'_i\,.
\end{equation}
Unlike neutron stars, in this case the amplitude of the photon field increases along the $z$ direction. The conversion length is
\begin{equation}
    L=\left(\dfrac{1}{2\pi k}\left\lvert\partial_z(\omega_p^2)\right\rvert\right)^{-1/2}=\sqrt{\dfrac{\pi k}{\omega_p\partial_z\omega_p}}\,,
    \label{eq:LWD}
\end{equation}
and the conversion probability
\begin{equation}
    p_{\rm WD}\simeq \dfrac{\pi\kappa^2\omega_p^3}{3k\partial_z\omega_p}\,,
    \label{eq:pWDratio}
\end{equation}
When radial trajectories $\partial_z\simeq \partial_r$ are assumed, we obtain the conversion probability in the main text.

\section{Propagation of photon after production}
\label{sec:photopropagation}

Here we give further details on the evolution of the photon field after it is produced from dark photons. Provided they do not travel in a direction almost within the conversion surface (as is the case for the radial trajectories that we assume), after production at $r_c$ photons propagate in the plasma barely affected by the back-conversion effect so long as $\kappa\ll 1$. As the plasma frequency varies, the photon energy is conserved while its momentum is determined by the dispersion relation. Far away from the resonant conversion region the propagation equation in the main text is no longer valid because the energy-momentum of photons and dark photons do not match. Instead, the photon field evolves according to
\begin{equation}
    \dfrac{d^2\pmb{A}}{dr^2}+k^2(r)\pmb{A}=0\,,
    \label{eq:photonpropagation}
\end{equation}
obtained by removing dark photon terms in Eq.~\eqref{eq:delAvec} and setting $B=0$. We write $\pmb{A}(r,t)=e^{i\omega t-ik(r)r}\tilde{\pmb{A}}(r)$ and assume the photon field varies slowly so that $|d^2\tilde{\pmb{A}}/dr^2|\ll k|d\tilde{\pmb{A}}/dr|$. Using the WKB approximation  Eq.~\eqref{eq:photonpropagation} simplifies to
\begin{equation}
    2k(r)\dfrac{d\tilde{\pmb{A}}}{dr}+\tilde{\pmb{A}} (r)\dfrac{dk}{dr}=0\,.
    \label{eq:photonpropagationsimplified}
\end{equation}
Eq.~\eqref{eq:photonpropagationsimplified} yields the relation $\tilde{\pmb{A}} (r)\propto 1/\sqrt{k(r)}$. In addition, the amplitude of the photon field will drop as a function of radius, analogously to the case of black body radiation. In combination, we find at a radius $r>r_c$,
\begin{equation}
    \tilde{\pmb{A}} (r)=\tilde{\pmb{A}} (r_c)\dfrac{r_c}{r}\sqrt{\dfrac{\mAp v_c}{k(r)}}\,,
    \label{eq:Apropagation}
\end{equation}
where as before $v_c$ is the velocity of dark photon at the resonant conversion radius. Eq.~\eqref{eq:Apropagation} simply shows the total photon flux, proportional to $|\tilde{\pmb{A}}|^2r^2k$, is conserved during the photon propagation, as in~\cite{Millar:2021gzs,Leroy:2019ghm}.

The photon propagation in an anisotropic plasma with a strong magnetic field (\eg~neutron stars) is more complicated, and in this case the propagation equation should be solved explicitly. Likewise the propagation in a non-magnetic cataclysmic variable will be complicated by the anisotropy of the plasma, which can lead to photons refracting despite the lack of a strong magnetic field. We leave a dedicated study for future work, and instead we assume Eq.~\eqref{eq:Apropagation} to hold also in these environments (which is reasonable given the interpretation as conservation of the total photon flux).

Assuming Eq.~\eqref{eq:Apropagation},  the photon signal power per unit solid angle outside the plasma is
\begin{equation}
    \dfrac{d\mathcal{P}}{d\Omega}=r^2|\pmb{S}|=\dfrac{1}{2}\omega^2 r^2|\pmb{A}|^2=\dfrac{1}{2}r_c^2\mAp\omega v_c|\pmb{\tilde{A}}(r_c)|^2\,,
\end{equation}
where the Poynting flux $\pmb{S}=\pmb{E}\times \pmb{H}$. The converted photon field $\pmb{\tilde{A}}(r_c)$ can be inferred from the conversion probability and the energy density of dark photon $\rho_{A'}=\dfrac{1}{2}\omega^2|\pmb{A}'|^2$, which yields
\begin{equation}
    \dfrac{d\mathcal{P}}{d\Omega}\simeq 2pr_c^2\rho_{A'}(r_c)v_c\,,
\end{equation}
where $p$ is the conversion probability given in Eq.~\eqref{eq:pNSratio} and~\eqref{eq:pWDratio} for neutron stars and white dwarfs, respectively. We have used the approximation $\omega\simeq \mAp$ and included a factor of 2 to account for dark photon conversion when entering and leaving the resonant conversion region.

\section{Description of radio telescopes}
\label{sec:radiotelescopes}

Here we briefly review the properties of radio telescopes, in particular the minimum detectable flux density, which we used to determine the sensitivity to dark photons in the main text. 

\begin{figure}
    \centering
    \includegraphics[width=0.6\textwidth]{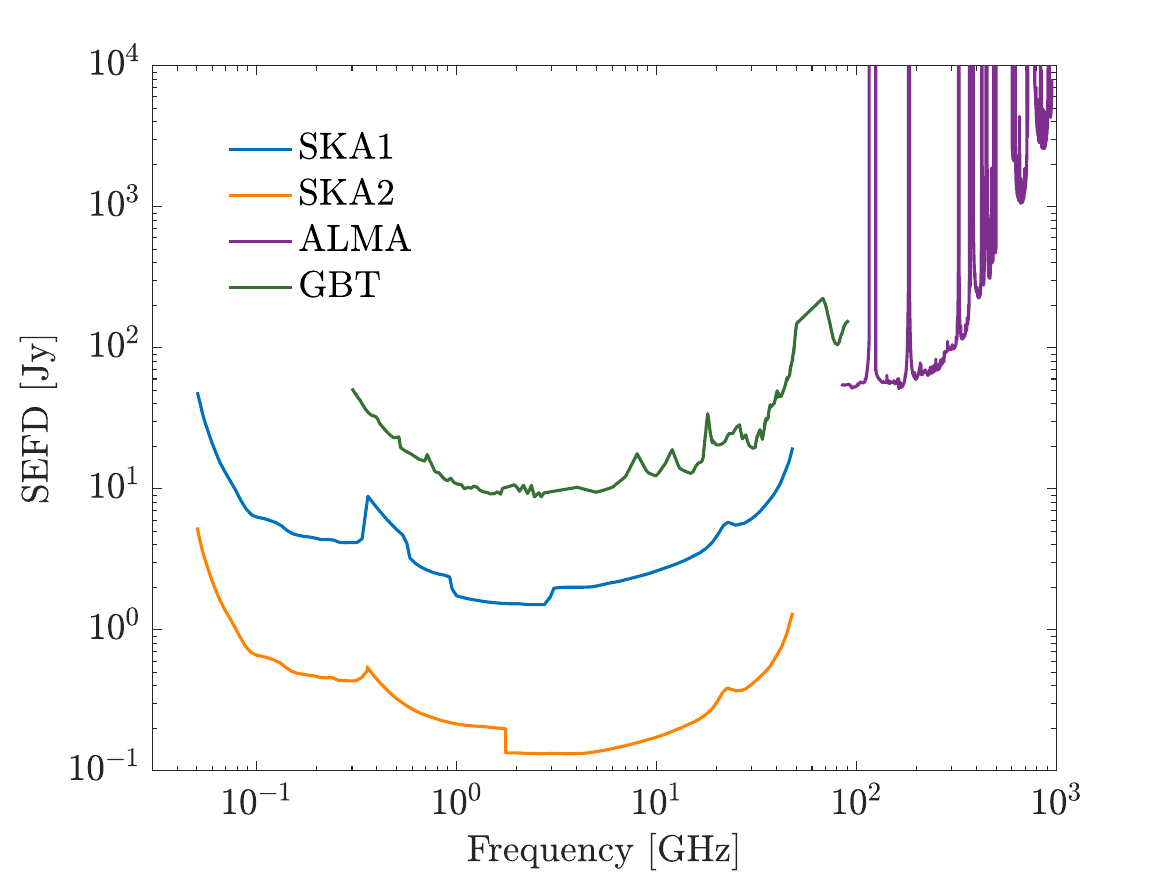}
    \caption{Configuration sensitivity, in particular the system equivalent flux density, of proposed future radio telescopes as a function of the signal frequency.} 
    \label{fig:radiotelescope}
\end{figure}

The minimum detectable flux density of a radio telescope is defined to be
\begin{equation}
    S_{\min}=\dfrac{\rm SEFD}{\eta\sqrt{n_{\rm pol}\mathcal{B}t_\mathrm{obs}}}\,,
    \label{eq:Smin}
\end{equation}
where $\eta$ is the detection efficiency, $t_\mathrm{obs}$ is the integrated observation time, and $n_{\rm pol}$ is the number of detected signal polarisations, which is 2 for most telescopes. SEFD is the system equivalent flux density,
\begin{equation}
    \mathrm{SEFD}=2k_B\dfrac{T_{\rm sys}}{A_{\rm eff}}=2.75~{\rm Jy}\dfrac{1000~\mathrm{m}^2/\mathrm{K}}{A_{\rm eff}/T_{\rm sys}}\,,
    \label{eq:SEFD}
\end{equation}
with $k_B$ the Boltzmann constant, $T_{\rm sys}$ the system temperature and $A_{\rm eff}$ the effective antenna area.
At frequencies below 1~GHz, there is a sizable radio background expected from synchrotron radiation in the galactic centre, however this is negligible at higher frequencies that we focus on~\cite{Safdi:2018oeu,GBT}. Additionally, at frequencies close to or above THz, absorption by the Earth's atmosphere is catastrophic for the detectable signal for Earth based telescopes (the effect of this is incorporated in $\rm{SEFD}$). We discuss the radio telescopes considered in this work below, and plot their sensitivities in Fig.~\ref{fig:radiotelescope}.

{\bf GBT.} We take the SEFD of GBT from Ref.~\cite{GBT} (with typical galactic background). 
We also assume the same detection bandwidth in SKA phase 2 as in phase 1 in similar frequency bands. The configuration sensitivity of SKA2 is roughly 15 times better than SKA1.  We take the detector efficiency $\eta=0.9$ for SKA1~\cite{dewdney2013ska1} and $\eta=1$ for SKA2 and GBT. The configurations of different radio telescopes are listed in Table~\ref{tab:radiotelescopes}.

{\bf SKA.} The configurations of SKA1 and SKA2 are obtained from~\cite{dewdney2013ska1}  and their sensitivities are computed in Ref.~\cite{Braun:2019gdo}.

\begin{table}[t!]
\centering
\setlength\extrarowheight{3pt}
\begin{tabular}{l | l | l }
\hline\hline
	Telescope & Band~[GHz] & $\mathcal{B}_{\rm det}$~[kHz] \\ \hline
    SKA1-LOW & [0.05, 0.35] & 1 \\
     SKA1-MID B1-B2& [0.35, 1.76] & 3.9  \\
     SKA1-MID B3-B5+& [1.65, 50] & 9.7 \\
     GBT & [0.1, 116] & 2.8 \\
     ALMA B3-B10& [84, 950] & 15.3 \\
\hline \hline
\end{tabular}
\caption{The frequency bands and minimum detection bandwidth of different radio telescopes.}
\label{tab:radiotelescopes}
\end{table}

{\bf ALMA.} For ALMA we follow the prescription in Ref.~\cite{ALMA}. The flux sensitivity for the 12-meter and 7-meter arrays is given by 
\begin{equation} \label{eq:Smin2}
    S_{\min}=\dfrac{2w_rk_B T_{\rm sys}}{\eta_q\eta_c(1-f_s)A_{\rm eff}\sqrt{N(N-1)n_{\rm pol}\mathcal{B}t_\mathrm{obs}}}\,,
\end{equation}
where $w_r=1.1$ is the weighting factor, $\eta_q=0.96$ is the quantization efficiency, $\eta_c=0.88$ is the correlator efficiency, and $N=50(12)$ for the number of antennas in the 12(7)-meter arrays. We take the shadowing fraction $f_s=0$. The effective area $A_{\rm eff}=A_{\rm an}\eta_{\rm ap}$. We focus on the 12-meter arrays with physical area $A_{\rm an}=113.1$~m$^2$ and the aperture efficiency $\eta_{\rm ap}$ ranges from 71\% to 31\% from Band 3 to Band 10. ALMA uses two types of correlators to identify fine spectral lines. We adopt the general analysis channel width of 15.3~kHz as suggested in the handbook, which is typically below the signal bandwidth of dark photons with masses $\gtrsim 10^{-4}$~eV (both for emission from a single white dwarf and also from a collection of neutron stars). We use the $T_{\rm sys}$ at the first octile of PWV (best weather conditions) in~\cite{ALMA}, and, for plotting in Fig.~\ref{fig:radiotelescope}, we show ${\rm SEFD}\equiv w_rT_{\rm sys}/(\eta_q\eta_c(1-f_s)NA_{\rm eff})$ for the 12-meter arrays. ALMA's sensitivity is lost at particular frequencies that correspond to  gaps between different bands or values where atmospheric absorption is significant. 


Finally, we note that the analysis bandwidth of the Breakthrough Listening (BL) project is 91.6~kHz, narrower than the width of the signal from neutron stars. Consequently, the bound on $S_{\rm{min}}$ that we impose to obtain the limit from neutron stars in the main text is stronger than the experimental published value by a factor of $\sqrt{\mAp\sigma_s/91.6~\rm kHz}$ due to Eq.~\eqref{eq:Smin}.

\section{The compact star population in the galactic centre} \label{sec:profiles}


Observations show there exist dense and luminous nuclear star clusters within the central few parsecs of the Milky Way galaxy, which consist of main sequence stars, stellar black holes and compact stars~\cite{genzel2010galactic}. Monte Carlo simulations~\cite{freitag2006stellar} suggest the compact star populations in the galactic centre can (assuming an average white dwarf mass of 0.6~$M_\odot$ and an average neutron star mass of 1.4~$M_\odot$) be fit by the relations
\begin{equation}
\begin{split}
    &n_{\rm NS}(R)=2.32\times 10^6 x^{-1.62}~{\rm pc}^{-3}\,,\\
    &n_{\rm WD}(R)=10^{-0.226 \log_{10}^2x-0.701 \log_{10}x+6.75}~{\rm pc}^{-3}\,, \\ 
\end{split} \label{eq:nwdnns}
\end{equation}
where $x=R/0.01$~pc. These relations are thought to be accurate for $R$ between 0.01~pc and 20~pc for white dwarfs, and from 0.025~pc to 8~pc for neutron stars.

The matter potential in the galactic centre region relevant to the signals we consider is dominated by the central black hole, which we assume to have a mass of $3.5\times 10^6M_\odot$. In addition, there are contributions from dark matter that depend on the density distribution assumed. The galactic circular velocity is $v_0=\sqrt{GM(R)/R}$, where $M(R)$ is the mass enclosed in radius $R$. To estimate the signal dispersion, we approximate $v_0\propto R^\alpha$ with $\alpha\simeq -1/2$, which is roughly correct for both dark matter profiles at $R<1$~pc, where the enclosed dark matter mass is less than the black hole.  We also approximate the number density of stars $n\propto R^{-\gamma}$. For neutron stars $\gamma\simeq 1.6$, and this numerical value also fits the white dwarf number density in Eq.~\eqref{eq:nwdnns}  reasonably well in the galactic centre. 
From the Jeans equation the velocity dispersion of stars is then $\sigma_s\simeq \sqrt{1/(\gamma-2\alpha)}v_0$~\cite{Dehnen:2006cm,Safdi:2018oeu}. As mentioned in the main text, when considering the cumulative signal from neutron stars we use the star velocity dispersion at $R=0.1$~pc. For the gNFW profile this yields $v_0=389$~km/s, $\sigma_s=240$~km/s and for a density spike $v_0=406$~km/s, $\sigma_s=251$~km/s. The central black hole is assumed to have a mass of $3.5\times 10^6M_\odot$. 

Finally, we consider the distribution of cataclysmic variables. Recently the X-ray point sources in the galactic centre have been revisited in~\cite{zhu2018ultradeep} based on Chandra observations~\cite{zhu2018ultradeep}. Due to the gravitational potential of the central black hole, frequent encounters of massive stars facilitate the formation of binaries and cataclysmic variables, which are believed to be the origin of such X-rays. The surface number density (number density per area) of observed $2-8$~keV X-ray point sources in the galactic centre can be well described by~\cite{zhu2018ultradeep}
\begin{equation}
    \log_{10}\dfrac{\Sigma}{\rm{arcsec}^2}=-0.0596y^3+0.00262y^2-0.188y-1.52\,,
\end{equation}
where $y=\log_{10}(R/\rm arcsec)$ and $R\leq 100''$. Given this number density, our assumption in the main text of at least one appropriately aligned cataclysmic variable within $0.3~{\rm pc}$ is reasonable.

\section{Radio signals from non-accreting white dwarf atmospheres and possible coronae} \label{app:WDnon}

Here we consider signals from dark photons converting in non-accreting white dwarfs. In particular, we analyse conversion in the atmosphere of white dwarfs and in the corona that might surround some white dwarfs.

\begin{figure}
    \centering
    \includegraphics[width=0.48\textwidth]{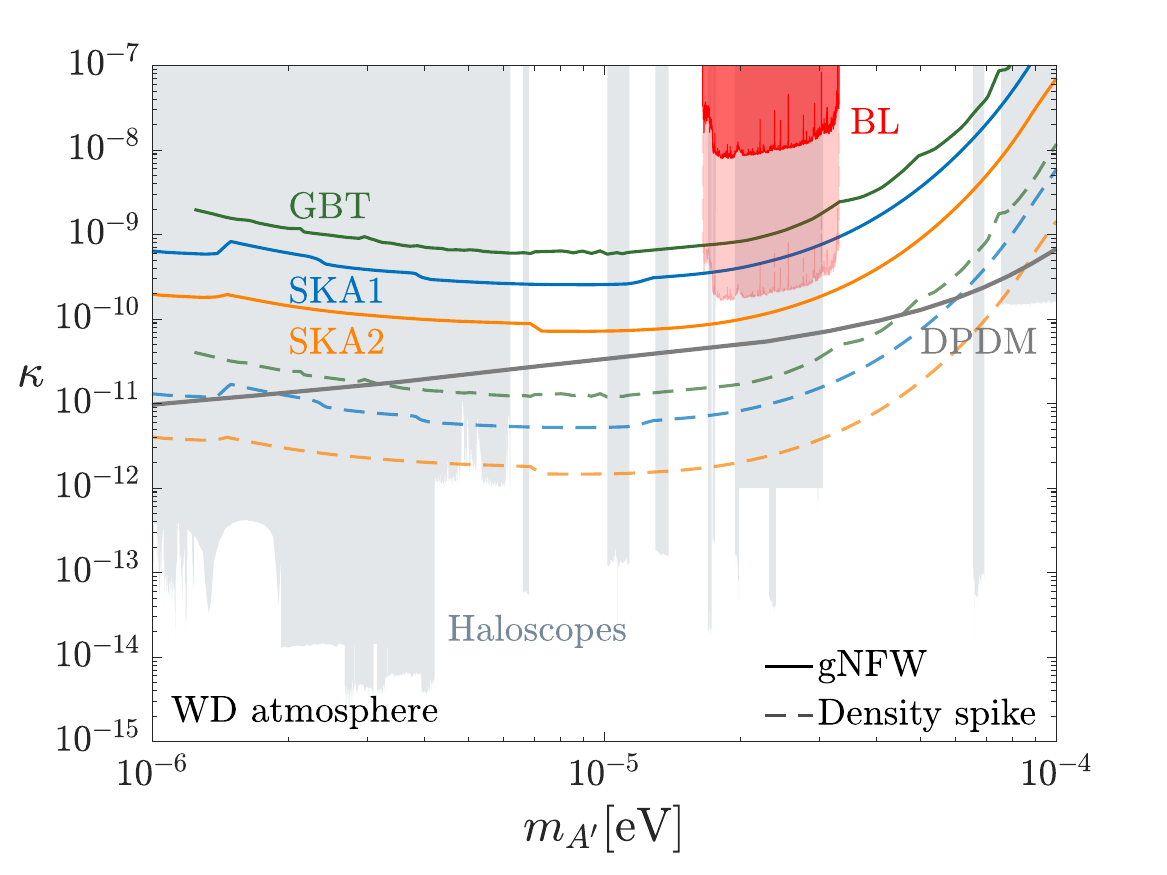}
    \includegraphics[width=0.48\textwidth]{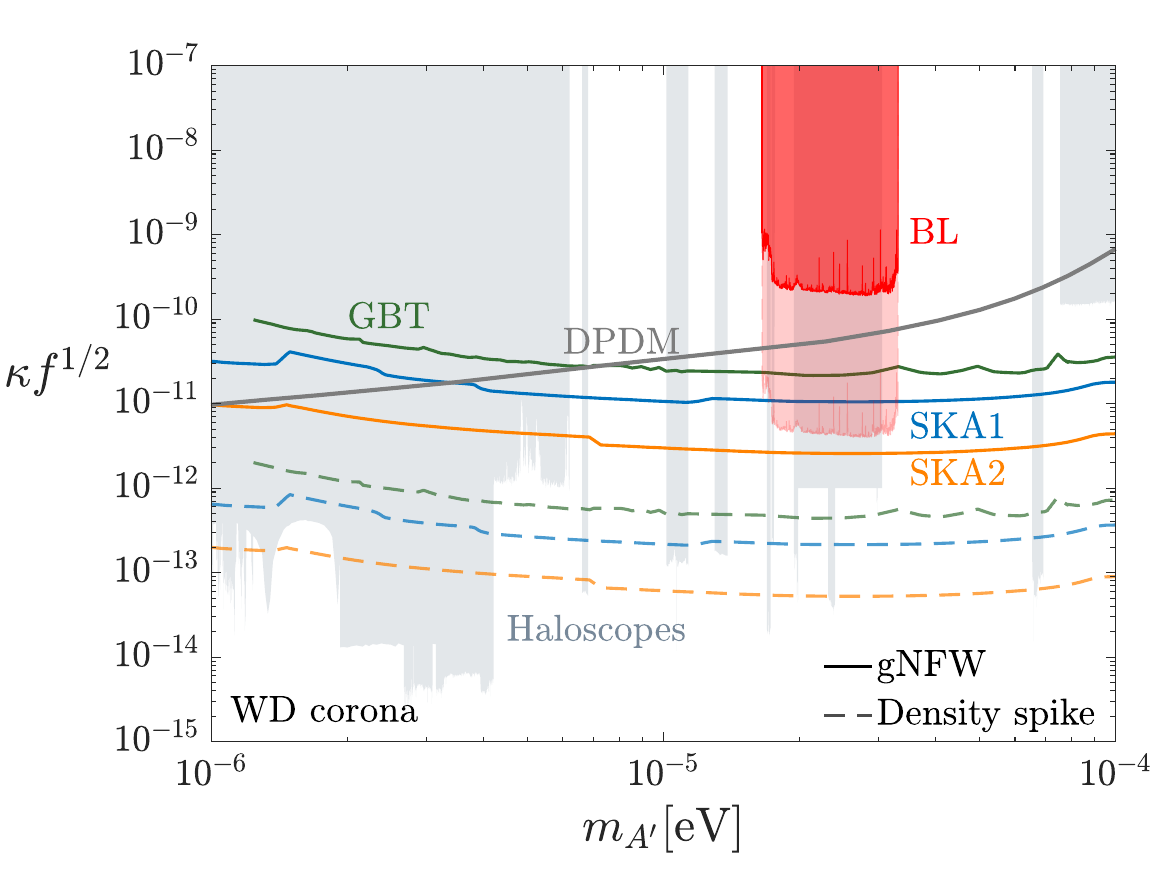}    
    \caption{Projected sensitivity of radio telescopes to dark photon dark matter from 100 hours of observation of the cumulative signals from the atmosphere (left) and the possible corona (right) of white dwarfs within 3~pc of the galactic centre. $f$ is the fraction of white dwarfs with a dense enough corona for resonant conversion, see text for details. As in the figure in the main text, coloured lines show the projected sensitivity from GBT, SKA1, and SKA2 and the red shaded regions show our constraints derived from the Breakthrough Listen (BL) project~\cite{Foster:2022fxn}; for each of these we show results both for a gNFW and a density spike profile.}    
    \label{fig:WDatm}
\end{figure}

White dwarfs typically have a hot, dense atmosphere with an effective temperature of up to $2\times 10^5$~K. DA type white dwarfs with hydrogen-dominated atmosphere make up the majority of the observed white dwarfs, and the temperature of these spans the range 4000~K to $1.2\times 10^{5}$~K~\cite{hoard2011white}. This motivates us to consider a log-normal distribution $f_T$ for the white dwarf atmosphere temperature centred at $\log_{10}(T_a/\rm K)$=4.34 and $\sigma_{\log_{10}(T_a/\rm K)}=0.4$. Because we expect the pressure gradient $\rho^{-1}\partial P/\partial r$ to balance the gravitational potential $GM_{\rm WD}/r$, we approximate the scale height of a WDs atmosphere
\begin{equation}
    l_a\simeq \dfrac{kT_a r_0^2}{GM_{\rm WD}\mu m_p}=0.06~{\rm km}\left(\dfrac{T_a}{10^4~\rm K}\right)\left(\dfrac{M_{\rm WD}}{M_\odot}\right)\left(\dfrac{r_0}{0.01~R_\odot}\right)^2\,,
\end{equation}
and the free electron density profile
\begin{equation}
    n_e(r)=n_0\exp\left(\dfrac{r-r_0}{l_a}\right)\,,
    \label{eq:neofr}
\end{equation}
where we take the mean molecular weight $\mu=0.5$ for fully ionized hydrogen plasma. Based on spectroscopic studies, we assume a maximum electron number density  $n_0=10^{17}$~cm$^{-3}$~\cite{kieu2017new,tremblay2009spectroscopic} (the same argument also applies to the boundary layer and is consistent with the description in~\cite{ingham1976origin,Gill:2011yp,Wang:2021wae}). Unlike the boundary layer around accreting white dwarfs, the atmosphere of a general white dwarf is expected to extend to a much larger radius than $r_0+l_a$. The magnetic field in white dwarfs is highly uncertain, but it is believed that only about 10\% of white dwarfs have a magnetic field stronger than 0.1~MG~\cite{Wang:2021hfb,kawka2007spectropolarimetric,holberg201625,hollands2015incidence} so we set $B=0$. The radio signals from the white dwarf atmosphere are therefore emitted isotropically. The signal power 
\begin{equation}
    \dfrac{d\mathcal{P}_{\rm WDa}}{d\Omega}
    =2.3\times 10^{7}~{\rm W}\ \left(\dfrac{\kappa}{10^{-8}}\right)^2
    \left(\dfrac{\mAp}{10^{-5}~\rm eV}\right)
    \left(\dfrac{\rho_{A'}^\infty}{0.3~{\rm GeV/cm}^3}\right)
\left(\dfrac{300~\rm km/s}{ v_0}\right)\left(\dfrac{T_a}{10^4~\rm K}\right) p_s^{\rm{IB}}\,.
\end{equation}
where we have fixed $M_{\rm WD}=M_\odot$ and $r_0=0.01R_\odot$. 
This is markedly lower than the signal from accreting white dwarfs, but the signal can be enhanced by considering a collection of stars. Similarly to neutron stars, the signal flux density from the galactic centre
\begin{equation}
        {S}_{\rm sig} = \dfrac{1}{\mathcal{B}d^2}\int_{R_{\min}}^{R_{\rm max}}4\pi R^2 n_{\rm WD}(R) dR
        \int f_T \dfrac{d\mathcal{P_{\rm WDa}}}{d\Omega} d\log_{10}T_a \,.
\label{eq:sigtotWDa}
\end{equation}
where we take $R_{\min}=0.01$~pc from simulations and $R_{\min}=3$~pc from the view of radio telescopes. We assume the average white dwarf mass of 0.6~$M_\odot$ and the corresponding radius of 0.012~$R_\odot$~\cite{yuasa2010white,nauenberg1972analytic}. The resulting sensitivity is shown in the left panel of Fig.~\ref{fig:WDatm}.

Additionally, it is possible that some white dwarfs might be surrounded by a hot envelope in the outer part of their atmosphere, a so-called ``corona'' analogous to the solar corona~\cite{fleming1993detection,arnaud1992coronal}. Such corona was originally proposed to account for observations of X-ray emission by telescopes in Earth orbits, including by Einstein, EXOSAT, ROAST and Chandra~\cite{musielak2003chandra}, which could be the result of plasma emission in such a region. However, these observations were later revisited and found to be either consistent with emissions
from the photosphere or with a non-detection~\cite{weisskopf2007chandra,musielak2003chandra,drake2005deposing}. Upper limits were set on the electron density in the corona, which range from $4.4\times 10^{11}$ to $5\times 10^{12}$~cm$^{-3}$ ~\cite{weisskopf2007chandra,zheleznyakov2004thermal}.

Nevertheless, it remains possible that corona could exist in some white dwarfs, so we briefly consider the radio signals that this would lead to. The suggested temperatures of corona $T_c$ range from $10^6$~K to $\gtrsim 10^7$~K~\cite{zheleznyakov2004thermal,weisskopf2007chandra}. We therefore assume a log-normal distribution with centre $\log_{10}(T_c/\rm K)$=6 and $\sigma_{\log_{10}(T_c/\rm K)}=1$. The form of the electron density profile would be similar to that in white dwarf atmospheres, and the signal power is given by Eq.~\eqref{eq:sigtotWDa}, albeit with a different temperature distribution. The signal flux density can be again computed from Eq.~\eqref{eq:sigtotWDa}. We remain agnostic about the maximum electron density in the corona $n_0$, but we include a factor $f$ that quantifies the fraction of white dwarfs with a corona that is dense enough that $n_0\geq \frac{m_e\mAp^2}{4\pi\alpha}$. Given the non-observation of corona emission, this fraction is likely to be small for $n_0\gtrsim 10^{12}~\rm{cm}^{-3}$. The sensitivity of radio telescopes to the relevant combination $\kappa f^{1/2}$ is plotted in the right panel of Fig.~\ref{fig:WDatm}.

\section{Attenuation of photons after production}
\label{sec:attenuation}

As the photons travel out of the white dwarf, the leading processes in the  plasma that might reduce the signal photon flux are Compton scattering and inverse bremsstrahlung. The former occurs at a rate $\Gamma_{\rm Comp}=8\pi\alpha^2n_e/(3m_e^2)$ and broadens the frequency range of the photon signal (we conservatively assume that the fraction of the signal that is scattered is lost). The latter leads to absorption of the photon flux and occurs with the rate~\cite{An:2020jmf,Redondo:2008aa}
\begin{equation}
    \Gamma_{\rm IB}=\dfrac{8\pi \alpha^3n_en_{\rm ion}}{3\omega^3m_e^2}\sqrt{\dfrac{2\pi m_e}{T}}\ln\left(\dfrac{2T^2}{\omega_p^2}\right)\left(1-e^{-\omega/T}\right)\,,
    \label{eq:GammaIB}
\end{equation}
where $n_{\rm ion}$ is the ion number density. We always assume a hydrogen-dominated plasma so $n_{\rm ion}\simeq n_e$.
Assuming photons travel on radial trajectories, the survival probability against scattering processes is
\begin{equation}
    p_s=\exp\left(-\int \Gamma dt\right)=\exp\left(-\int_{r_c}^{\infty}\Gamma/v_\gamma dr\right)\,.
\end{equation}
In a non-magnetised plasma photons travel at velocity $v_\gamma=\sqrt{1-\omega_p^2/\omega^2}$. Notice that both $\Gamma$ and $v_\gamma$ depend on the electron density profile, which may vary depending on the environments. We assume an exponential density profile as described in Eq.~\eqref{eq:neofr}, which is expected to be valid for the white dwarf atmosphere, corona or boundary layer plasmas. Evaluating the resulting integral gives
\begin{equation}
    p^{\rm Comp}_s=\exp\left(-\dfrac{2\alpha l_p \mAp^2}{3m_e}\right)\,,
\end{equation}
and
\begin{equation}
    p^{\rm IB}_s=\exp\left(-\dfrac{2\alpha l_p\mAp}{27\pi}\left(\dfrac{2\pi m_e}{T_p}\right)^{1/2} \right.
    \left.\left(1-e^{-\mAp/T_p}\right)\left(3\ln\left(\dfrac{2T_p^2}{\mAp^2}\right)+0.84\right)\right)\,,
\end{equation}
where $l_p$ is the scale height of the plasma, $T_p$ is the temperature of the plasma (which we assume to be a constant), and we approximated $\omega\simeq \mAp$. The attenuation due to Compton scattering is negligible in the dark photon mass range we consider. Although Eq.~\eqref{eq:GammaIB} suggests that the inverse bremsstrahlung rate diverges as $\omega\sim\mAp\to 0$, the survival probability actually decreases quickly as the dark photon mass increases due to the $n_e^2\propto \omega_p^4\sim \mAp^4$ term in the numerator (this is because dark photons with larger mass convert at locations with larger plasma density, where absorption is more efficient). 

Photons produced in a boundary layer travel from $r_c$ to $r_0+b$ approximately radially. Depending on the dark photon mass and height above the disk, this distance ranges from 0 to $b$. The photons will be slightly deflected in the direction of the density gradient, where less dense plasma is expected than the exponential radial density profile. Although we do not carry out detailed modelling, to estimate the possible effect in accreting white dwarfs, considered in the main text, we include attenuation assuming all photons travel radially over a distance of 500~km, which yields \begin{equation}
    p^{\rm IB}_{s,\rm CV}=\exp\left(-\dfrac{2\alpha b\mAp}{27\pi}\left(\dfrac{2\pi m_e}{T_s}\right)^{1/2} \right.
    \left.\left(1-e^{-\mAp/T_s}\right)\left(1.63\ln\left(\dfrac{2T_s^2}{\mAp^2}\right)+0.079\right)\right)\,,
\end{equation} 
for $b=3188$~km, as adopted in the main text, where $T_s$ is the boundary layer temperature. Note that the coefficients in the last bracket depend implicitly on $b$ and the distance of travel.

The propagation of photons in neutron stars is even more complex and a reliable prediction requires ray tracing. However, for the purpose of estimating the attenuation of signals in the magnetosphere of neutron stars analysed in the main text we continue to assume radial photon trajectories and a dispersion relation $\omega^2=k^2+\omega_p^2$. Using Goldreich--Julian (GJ) model~\cite{Goldreich:1969sb}, the resultant survival probability due to Compton scattering and inverse bremsstrahlung is
\begin{equation}
    p^{\rm Comp}_{s,\rm NS}=\exp\left(-0.57\dfrac{\alpha r_c \mAp^2}{m_e}\right)\,,
\end{equation}
and
\begin{equation}
    p^{\rm IB}_{s,\rm NS}=\exp\left(-\alpha r_c\mAp\left(\dfrac{2\pi m_e}{T_m}\right)^{1/2} \right.
    \left.\left(1-e^{-\mAp/T_m}\right)\left(0.078\ln\left(\dfrac{2T_m^2}{\mAp^2}\right)+0.027\right)\right)\,,
\end{equation}
where $r_c$ is the radius of the conversion region and $T_m$ is the temperature of the magnetosphere, which ranges from $\simeq 10^6$ to $\simeq 5\times 10^6$~K~\cite{mori2007modelling}. The Compton scattering is also very inefficient in the magnetosphere and it seems likely this will remain the case after a full analysis. 

\begin{figure}[!h]
    \centering
    \includegraphics[width=0.48\textwidth]{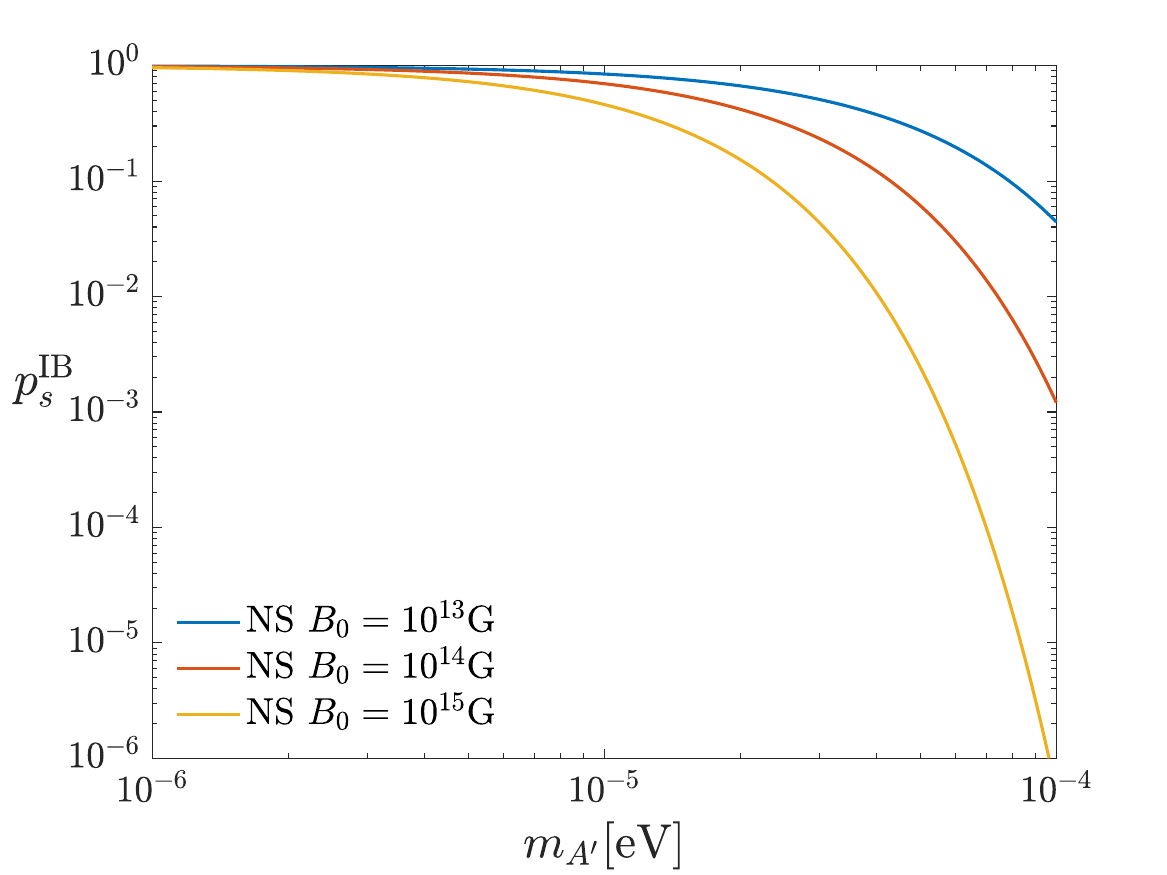}
    \includegraphics[width=0.48\textwidth]{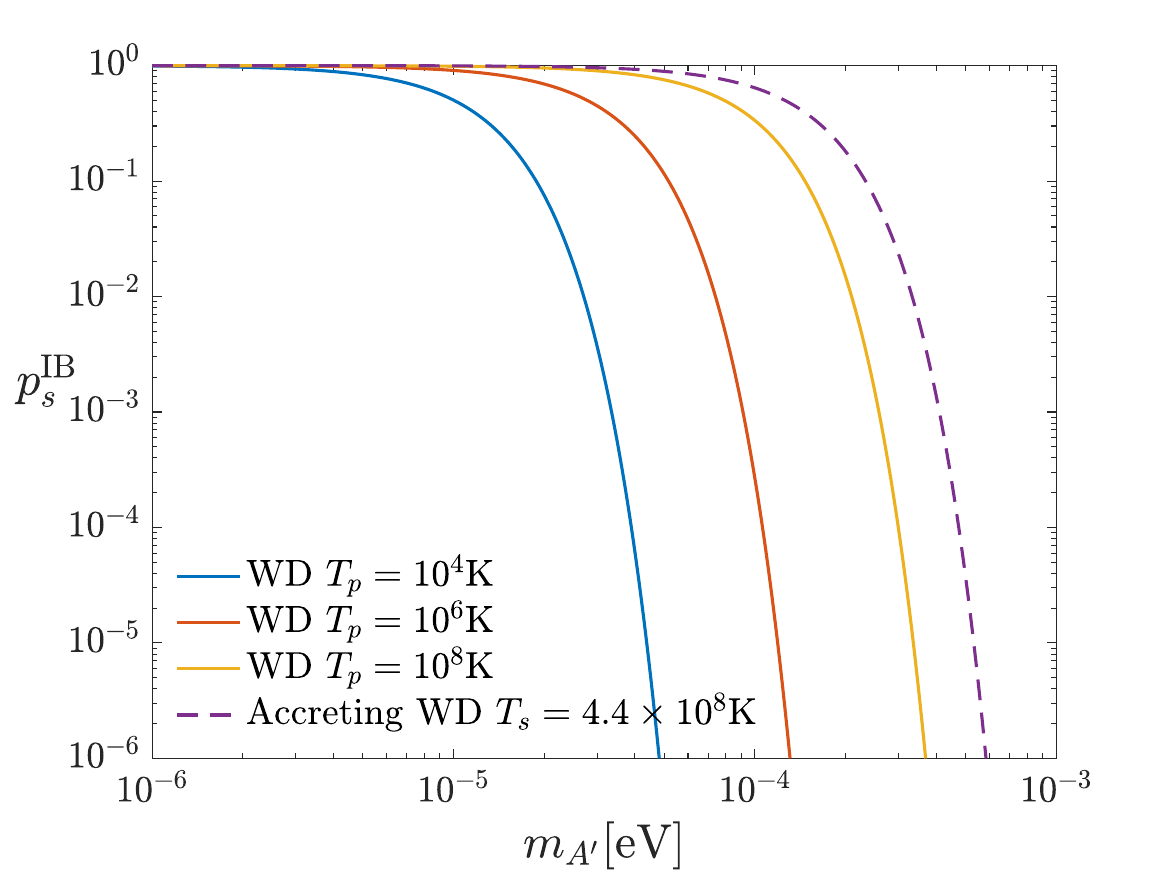}    
    \caption{The survival probability of the converted signal photons due to inverse bremmsstrahlung in the plasma. {\it Left:} The survival probability in the neutron star magnetosphere. Different colors correspond to the surface magnetic field of $10^{13}$~G, $10^{14}$~G and $10^{15}$~G respectively. We assume the spin period $P=0.5$~s, the plasma temperature $T_m=2\times 10^6$~K and the angular dependence $\beta=1$. {Right:} The solid lines depict the survival probability of photons in the atmosphere of isolated white dwarfs, with the plasma temperature of $10^4$~K, $10^6$~K and $10^8$~K respectively. The dashed line delineates the absorption in non-magnetic cataclysmic variable with the boundary layer temperature $T_s=4.4\times 10^8$~K.}
    \label{fig:attenuation}
\end{figure}

We show the survival probability of signal photons in Fig.~\ref{fig:attenuation}. For neutron stars we assume $T_m=2\times 10^6$~K. As shown in the main text, the conversion radius $r_c$ is determined by the neutron star magnetic field, spin period and the emission angle. For higher magnetic field the absorption is more significant because $r_c$ is larger.  We find that the effect of inverse bremsstrahlung is unimportant unless $\mAp\sim 10^{-4}$~eV and $B_0\sim 10^{15}$~G. On the other hand, the absorption in the white dwarf plasma could be important depending on the plasma temperature.

\section{Dephasing and non-radial trajectories during conversion}
\label{sec:dephasing}

Here we study the impact of the dark photon trajectories not being exactly radial, as we have assumed so far. Such deviations can, for example, lead to `dephasing' of the photon and dark photon within the conversion length affecting the conversion probability.

\begin{figure}
    \centering
    \includegraphics[width=0.48\textwidth]{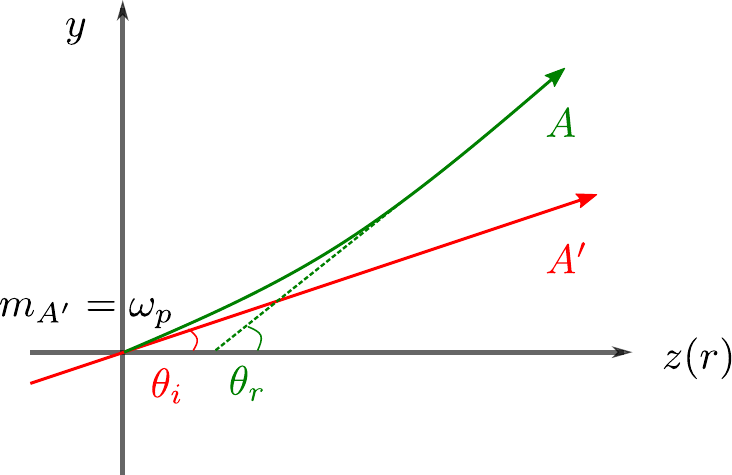}
    \includegraphics[width=0.48\textwidth]{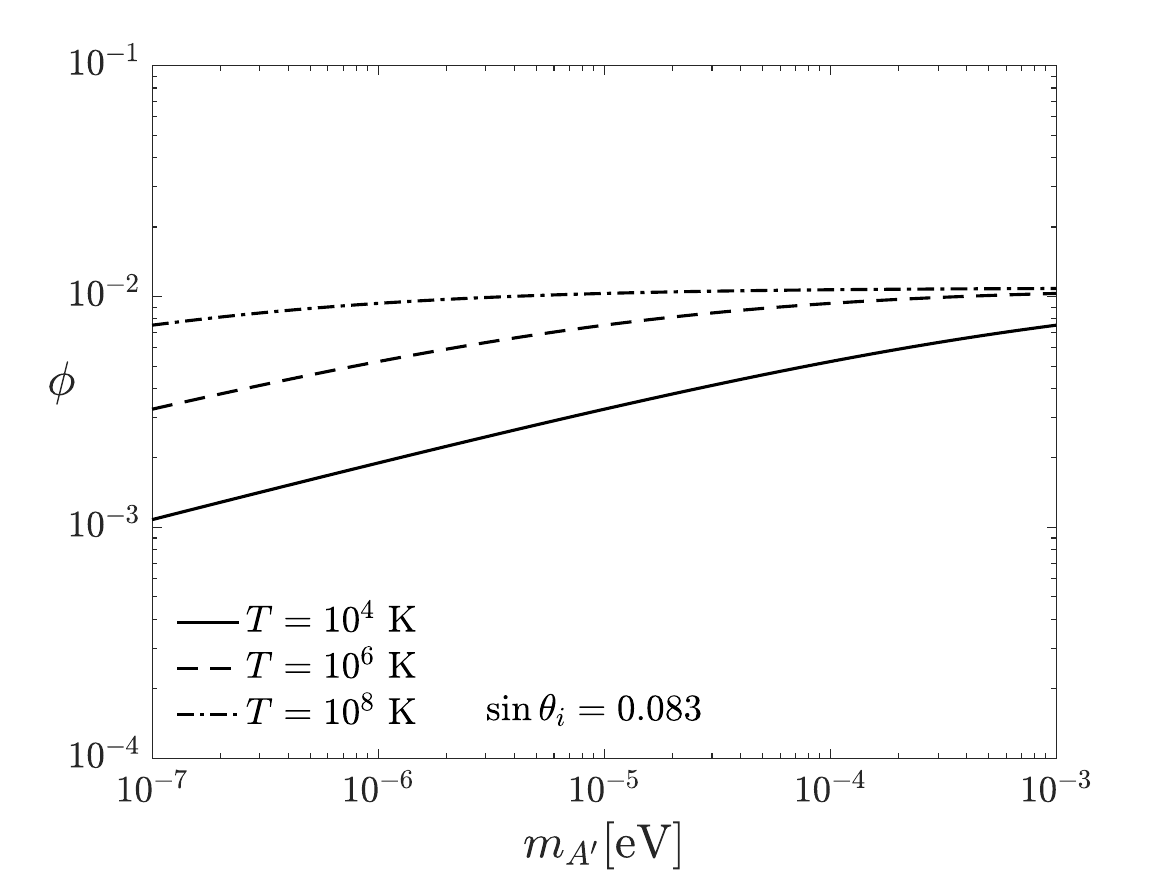}
    \caption{{\it Left:} Schematic illustration of photon and dark photon propagation in the conversion region. The $z$-axis is directed to the radial direction. Resonant conversion takes place at $z=0(r=r_c)$. {\it Right:} The dephasing factor $\phi$ as a function of dark photon mass in white dwarf environments with different properties. Different line styles correspond to plasma temperatures of $10^4$~K, $10^6$~K, and $10^8$~K, respectively.  We fix an incident angle $\sin\theta_i=0.083$ at the conversion region.}
    \label{fig:dephasing}
\end{figure}

We begin our analysis by assuming an isotropic environment, which is appropriate for the white dwarf atmospheres and coronae considered above. In this case deviations from radial trajectories occur due to the virial motion of the dark matter dark photons, which results in them entering the conversion zone, at radius $r_c$, with a small incident angle after being gravitationally accelerated. 
The propagation of dark photons and photons in such a conversion region is illustrated in Fig.~\ref{fig:dephasing}. We align the $z$-axis along the radial direction and choose the $y-z$ plane to be the plane of dark photon propagation. Given an initial halo dark photon velocity $v_i$, after gravitational acceleration at $z=0$, 
\begin{equation}
    \sin\theta_i=\dfrac{v_{i,y}}{\sqrt{v_i^2+v_c^2}}\simeq \dfrac{v_{i,y}}{v_c}\,.
\end{equation}
As a consequence of the non-radial trajectory, the phase integral in Eq.~\eqref{eq:exponentialphase} is modified to
\begin{equation}
    f(s')= -\int ds\dfrac{\mAp^2-\omega_p(s'')^2}{2k}\,,
    \label{eq:exponentialphases}
\end{equation}
where $s$ is interpreted as the trajectory of dark photon. Ignoring the gravitational curvature of the dark photon trajectory, which is small in the conversion region, the dark photon displacement $ds=dr/\cos\theta_i$. We can again expand Eq.~\eqref{eq:exponentialphases} to the second order, which gives the conversion length
\begin{equation}
    L\equiv \left(\dfrac{1}{2\pi k}\left\lvert\dfrac{d\omega_p(r')^2}{dr'}\cos\theta_i\right\rvert\right)^{-1/2}_{r'=r_c}\,.
    \label{eq:Lnonradial}
\end{equation}
For the typical halo dark matter velocity in the galactic centre $v_i\lesssim 400$~km/s and the escape velocity at a white dwarf surface $v_c\sim 4800$~km/s, $\sin\theta_i\lesssim 0.083$. Comparing Eq.~\eqref{eq:Lnonradial} with Eq.~\eqref{eq:LWD}, this introduces at most 0.17\% enhancement of the conversion length, which can be safely neglected.

Additionally, Eq.~\eqref{eq:exponentialphases} continues to assume straight-line trajectories for both the photon and dark photon. However, if either trajectory deviates from this assumption in the conversion region the conversion probability is further modified. In particular this effect, known as  \emph{dephasing}~\cite{Witte:2021arp}, reduces the phase overlap. It typically occurs when photons are refracted while propagating in the plasma. Although a more sophisticated modification of the field equations, along with ray-tracing, would be required to properly account for it, we estimate the impact of dephasing by investigating the integral
\begin{equation}
    \phi=\int  k_\gamma\left(\dfrac{dr}{\cos\theta_i}-\dfrac{dr}{\cos\theta_r}\right)\,,
\end{equation}
in the conversion region, where $\sin\theta_r\equiv k_{\gamma,y}/k_\gamma$. This quantifies the relative phase induced due to the difference in the paths of the photon and dark photon. We stress that this is robust against the choice of the (somewhat arbitrary) definition of the location of the conversion region relative to the resonant conversion surface. If the dark photon and photon both follow the same trajectory, this integral vanishes. The dephasing effect would be important if $|\phi|\sim 1$. Because $\phi$ appears in the exponential, its effect is similar to that of the non-radial trajectory, and we expect that the effect of dephasing can be parameterised by a modified conversion length $L'=L\sqrt{\cos\phi}$, with $L$ given in Eq.~\eqref{eq:Lnonradial}.  Without loss of generality we set $z=0$ to be at $r=r_c$. 
Snell's law holds in the plasma,
\begin{equation}
    n(r)\sin\theta_r=n_i\sin\theta_i\,,
    \label{eq:Snell}
\end{equation}
where $n$ is the refraction index and we define $n_i\equiv n(z=0)$. 
Indeed, in a plasma without a strong magnetic field the dispersion relation is $\omega^2=k_\gamma^2+\omega_p^2$, $n\equiv k_\gamma/\omega=(1-\omega_p/\omega)^{1/2}$, and we have the relation $n\sin\theta=k_{\gamma,y}/\omega={\rm const}$ as $dk_{\gamma,y}/dt=-\partial_y \omega=-\omega_p(\partial_y \omega_p/\omega)=0$ for a plasma frequency that changes only in the $z$ direction. Alternatively, $\theta_r$ can be obtained from $\sin\theta_r=k_{\gamma,y}/\sqrt{\omega^2-\omega_p^2}$, with $k_{\gamma,y}=k_{A'}\sin\theta_i$.
With this setup, we obtain
\begin{equation}
    \phi=\int_{r_c}^{r_L}n\omega dr\left[(1-w^2\sin^2\theta_i)^{-1/2}-\cos^{-1}\theta_i\right]\,,
    \label{eq:phiintWD}
\end{equation}
where $r_L=r_c+L\cos\theta_i$, and $w=n_i/n(r)$. At $z=0$, we have $\omega_p=\mAp$, and $\omega\simeq \mAp(1+v_c^2/2)$, which yields $n_i=[1-1/(1+v_c^2/2)^2]^{1/2}\simeq 0.016$. We note that the choice of limits in Eq.~\eqref{eq:phiintWD} corresponds to the conversion region being between $r=r_c$ and $r_c-L$, but our results are not sensitive to this choice. 
In Fig.~\ref{fig:dephasing} we plot the resulting dephasing factor $\phi$, computed from Eq.~\eqref{eq:phiintWD}, as a function of dark photon mass and for different plasma temperatures assuming a typical $\sin\theta_i$ (for smaller incident angles the dephasing is even less pronounced as the photon is refracted less). Despite our analysis being rough, the small values of $\phi$ obtained suggest that dephasing does not have a major effect in isotropic environments for typical incidence angles.


It is important to note that  Eq.~\eqref{eq:phiintWD} applies to isotropic plasma such as those in the white dwarf atmosphere or possible corona. The boundary layer in accreting white dwarf is slightly anisotropic with variations in the transverse direction. In such a scenario, the angles in Eq.~\eqref{eq:phiintWD} are to be defined with respect to the direction where the plasma density varies, and Eq.~\eqref{eq:phiintWD} should be modified accordingly. However, because the dephasing factor $\phi\ll 1$ in the isotropic plasma, we do not expect dephasing to dramatically modify the conversion probability in a boundary layer either. A full analysis, which would require reliable modeling of the boundary layer, is left for future work. Additionally, we have not investigated the dephasing in the neutron star magnetosphere, where a full 3-D ray-tracing is required to calculate the dephasing factor. Dephasing is thought to lead to significant corrections  for the case of axion conversion in this environment~\cite{Witte:2021arp}, and we expect this to also be true for dark photons.

\end{document}